\documentclass[useAMS,usenatbib]{mn2e}

\usepackage{graphicx}
\usepackage{amssymb}
\defcitealias{Lane09}{Paper I}

\newcommand{\new}[1]{{\rm #1}}

\title[Testing Newtonian Gravity with Globular Clusters]{Testing
  Newtonian   Gravity with AAOmega: Mass-To-Light Profiles and
  Metallicity   Calibrations From 47 Tuc and M55}
  \author[Richard R.  Lane {\it et al.}]{Richard R.
  Lane${^1}$\thanks{E-mail: rlane@physics.usyd.edu.au}, L\'aszl\'o L.
  Kiss${^{1,2}}$, Geraint F.  Lewis${^1}$, Rodrigo
  A. Ibata${^3}$,\vspace{2.5mm}\\
\hspace{-1mm}{\LARGE {\rm Arnaud Siebert$^{3}$,  Timothy
    R. Bedding${^1}$ and P\'eter Sz\'ekely$^{4}$}}\\
$^{1}$Sydney Institute for Astronomy, School of Physics, A29,
  University of Sydney, NSW, Australia 2006\\
$^{2}$Konkoly Observatory of the Hungarian Academy of Sciences, PO Box
    67, H-1525, Budapest, Hungary\\
$^{3}$Observatoire Astronomique, Universite de Strasbourg, CNRS,
  67000 Strasbourg, France\\
$^{4}$Department of Experimental Physics, University of Szeged, Szeged 6720,
Hungary\\}
\begin{document}

\date{This version: \today}

\pagerange{\pageref{firstpage}--\pageref{lastpage}} \pubyear{2002}

\maketitle

\label{firstpage}

\begin{abstract}

\noindent Globular  clusters are an  important test bed  for Newtonian
gravity  in  the  weak-acceleration  regime,  which is  vital  to  our
understanding of the nature  of the gravitational interaction.  Recent
claims  have  been  made  that  the velocity  dispersion  profiles  of
globular  clusters flatten  out at  large radii,  despite  an apparent
paucity  of dark matter  in such  objects, indicating  the need  for a
modification of gravitational theories.  We continue our investigation
of this claim,  with the largest spectral samples  ever obtained of 47
Tucanae  and  M55.   Furthermore,  this  large sample  allows  for  an
accurate metallicity calibration based on the equivalent widths of the
calcium triplet  lines and $K$  band magnitude of  the Tip of  the Red
Giant Branch.   Assuming an isothermal distribution,  the rotations of
each  cluster are also  measured with  both clusters  exhibiting clear
rotation signatures.   The global velocity dispersions of  NGC 121 and
Kron 3, two globular clusters  in the Small Magellanic Cloud, are also
calculated.   By applying  a simple  dynamical model  to  the velocity
dispersion   profiles  of  47   Tuc  and   M55,  we   calculate  their
mass-to-light profiles, total masses and central velocity dispersions.
We  find  no  statistically  significant flattening  of  the  velocity
dispersion at large radii for M55,  and a marked {\it increase} in the
profile of 47 Tuc for  radii greater than approximately half the tidal
radius.   We  interpret this  increase  as  an evaporation  signature,
indicating that 47 Tuc is undergoing, or has undergone, core-collapse,
but  find  no  requirement  for  dark  matter  or  a  modification  of
gravitational theories in either cluster.

\end{abstract}


\section{Introduction}

The  nature  of the  gravitational  interaction  is  one of  the  most
important concepts in astrophysics, yet complete comprehension of this
interaction  is  still  elusive.   The so-called  Pioneer  and  Fly-by
anomalies, where spacecraft exhibit  behaviour that is unexpected from
Newtonian  and general relativity  gravitation theories,  outline this
lack       of       understanding      \cite[see][and       references
therein]{Anderson02,deDiego08}, although these  examples may have more
mundane  explanations.  More  importantly,  it has  been claimed  that
several globular  clusters (GCs; $\omega$~Centauri, M15,  M30, M92 and
M107) exhibit  a flattening of  their velocity dispersion  profiles at
radii  $R\sim\frac{r_t}{2}$, where $r_t$  is the  tidal radius  of the
cluster \citep{Scarpa03,Scarpa04a,Scarpa04b}.   The authors claim that
either dark matter (DM), or a modification of gravitational theory, is
required to explain their results.

Modified theories of gravity  \cite[MOG; see][for a review of modified
gravity   theories]{Durrer08}   and   those  of   Newtonian   dynamics
\citep[MOND;][]{Milgrom83}  have  been  shown  to solve  some  of  the
discrepancies. However,  these are not universal  theories and to-date
have only  been applied to  specific instances \cite[e.g.   the Bullet
Cluster          and          galaxy         rotation          curves;
see][respectively]{Angus06,Sanders07}.   MOG   theories  diverge  from
Newtonian gravity  in the  high-acceleration regime and  MOND diverges
from Newton  in the low-acceleration  regime. Therefore, if  either of
these theories  were correct, the  effect should be measurable  at the
predicted accelerations.  Independent of MOG or MOND theories, testing
the gravitational interaction at low accelerations is essential to the
overall understanding of gravity.

Globular  clusters   are  an  ideal  testing   ground  for  weak-field
gravitation because  the accelerations  experienced by stars  at large
radii are below the limit  where DM, or a modified gravitation theory,
is  required  to  explain   observations  in  many  dynamical  systems
\cite[$a_0\approx1.2\times10^{-10}$\,m\,s$^{-2}$;][]{Scarpa07}.
Furthermore, they are thought to contain little, or no, dark matter --
indicated   by  dynamical   models  \citep{Phinney93},   {\it  N}-body
simulations   \citep{Moore96},   observations   of  GC   tidal   tails
\citep{Odenkirchen01},   dynamical   and   luminous  masses   of   GCs
\citep{Mandushev91} and  the lack of microlensing  events from GC-mass
dark  haloes \citep{Navarro97,Ibata02}, although  this is  still under
debate.  GCs are also located  at varying distances from the centre of
the  Galaxy,  so  that  if  all exhibit  similar  behaviour,  Galactic
influences cannot be the primary cause.

In \citealt{Lane09} (hereafter  \citetalias{Lane09}) we calculated the
velocity dispersions  and mass-to-light profiles of M22,  M30, M53 and
M68.   Our   conclusions  were  that  there  is   no  requirement  for
significant quantities of dark  matter, or a modification of Newtonian
gravity, to explain  the kinematics of any of  these clusters.  In the
current  paper  we  continue   this  investigation  with  the  largest
spectroscopic  dataset to date  of 47  Tucanae and  M55.  We  begin by
describing the data acquisition/reduction (Section \ref{data}) and the
membership selection for each cluster (Section \ref{membership}).  Our
large samples allow us to determine a metallicity calibration based on
the Tip  of the Red  Giant Branch (TRGB; Section  \ref{metaldist}), as
well  as the rotations  (Section \ref{rotation}),  systemic velocities
and    velocity   dispersions    (Section    \ref{dispersions}),   and
mass-to-light  profiles  (Section \ref{ML})  --  where  we also  place
limits  on  the  DM  content  of  each  cluster  from  their  velocity
dispersions  and  mass-to-light  profiles.   Finally,  our  concluding
remarks are presented in Section \ref{conclusions}.

\section{Data Acquisition and Reduction}\label{data}

AAOmega,  a   double-beam,  multi-object  spectrograph   on  the  3.9m
Anglo-Australian Telescope  (AAT) at Siding Spring  Observatory in New
South  Wales, Australia,  was employed  to  obtain the  data for  this
survey.  AAOmega  is capable of  obtaining spectra for  392 individual
objects over  a two degree field  of view. We used  the D1700 grating,
which  has been  optimized  for  the Ca  II  infrared triplet  region,
centered  on  8570\AA,  with  30  sky  fibres  used  for  optimal  sky
subtraction, and 5--8 fibres  for guiding.  The positional information
for  our targets  was  taken  from the  2MASS  Point Source  Catalogue
\citep{Skrutskie06}  which  has an  accuracy  of  $\sim0.1''$, and  we
selected stars that  matched the $J-K$ colour and  $K$ magnitude range
of the red giant branch (RGB) of each cluster.

Our observations were  performed over 7 nights on  August 12--18 2006,
and a  further 8 nights  on August 30  -- September 6 2007.   The mean
seeing was  $\sim1.5''$.  Several fibre configurations  were taken for
each cluster with 3600--5400 second exposures giving a signal-to-noise
of  $\sim50-250$.   To  minimize  scattered light  cross-talk  between
fibres, each field configuration was limited to stars in a 3 magnitude
range. In total, 4670 and 7462 spectra were obtained in the 47 Tuc and
M55  regions, respectively.  Flat  field and  arc lamp  exposures were
used    to   ensure   accurate    data   reduction    and   wavelength
calibration. \new{The pointing accuracy  of the AAT is $\sim0.3''$ and
the fibres have a $\sim2''$  diameter.  The offset due to the pointing
uncertainty is azimuthally scrambled by the fibre, so has no effect on
the  zero point of  the wavelength  calibration.}  Data  reduction was
performed               with               the              {\tt2dfdr}
pipeline\footnote{http://www2.aao.gov.au/twiki/bin/view/Main/CookBook2dfdr},
which is specifically  designed for AAOmega data. The  efficacy of the
pipeline  has been  checked with  a comparison  of  individual stellar
spectra.

Radial velocity  and atmospheric  parameters were obtained  through an
iterative  process which  takes the  best $\chi^2$  fits  to synthetic
spectra from the library  by \cite{Munari05} and cross-correlates this
model with  the observed spectra  to calculate the radial  velocity [a
process very  similar to that  used by the Radial  Velocity Experiment
\cite[RAVE;][]{Steinmetz06,Zwitter08}  project].   We  used  the  same
spectral  library  as the  RAVE  studies;  \cite{Kiss07} outline  this
process in detail.

\subsection{Radial Velocity and Uncertainty Estimates}\label{uncert}

To be sure that we are not under/overestimating the uncertainties, and
that  our  radial velocity  measurements,  and  \new{estimates of  the
random}  uncertainties,   are  reproducible,  we   observed  a  single
configuration   of   M68  (as   part   of   the   data  obtained   for
\citetalias{Lane09}), consisting of $317$ spectra of the same stars on
consecutive nights.

Radial  velocities from  the data  for a  single night  were estimated
using  two independent  pipelines  to  test the  efficacy  of our  own
software.   For this  we  compared  the outputs  of  our own  pipeline
\cite[again see][for a detailed description]{Kiss07} with that from the
pipeline  written  specifically for  the  RAVE  project.  The  average
difference  in   radial  velocities   between  the  two   pipelines  is
$0.3\pm0.1$\,km\,s$^{-1}$.

To  test  that  our   pipeline  reproduces  reliable  velocities,  and
associated  uncertainties, these  were  extracted from  the data  from
consecutive  nights.  Subtracting  one  from the  other  results in  a
distribution    with    a    mean   of    $-0.33$\,km\,s$^{-1}$    and
$\sigma\approx0.97$.   Therefore  95\%  of  the  stars  observed  have
velocities within $\approx2$\,km\,s$^{-1}$ from one night to the next.
This  is  within  the  systematics  of  the  instrument  (see  Section
\ref{dispersions}) and  is comparable to the  quoted uncertainties for
individual velocity estimates.  \new{Furthermore, because we used many
fibre configurations  for each cluster,  it was necessary to  test for
systematic  offsets between  configurations. Therefore,  we calculated
the  velocity dispersion  of stars  from four  separate configurations
within  the same  distance  bin  (13 stars  were  available from  each
configuration). The  dispersions between configurations  had a maximal
difference   of   $0.3$\,km\,s$^{-1}$,   which   is  well   within   the
uncertainties of the bin.}

\subsection{Cluster Membership}\label{membership}

\begin{figure*}
  \begin{centering}
  \includegraphics[angle=-90,width=0.45\textwidth]{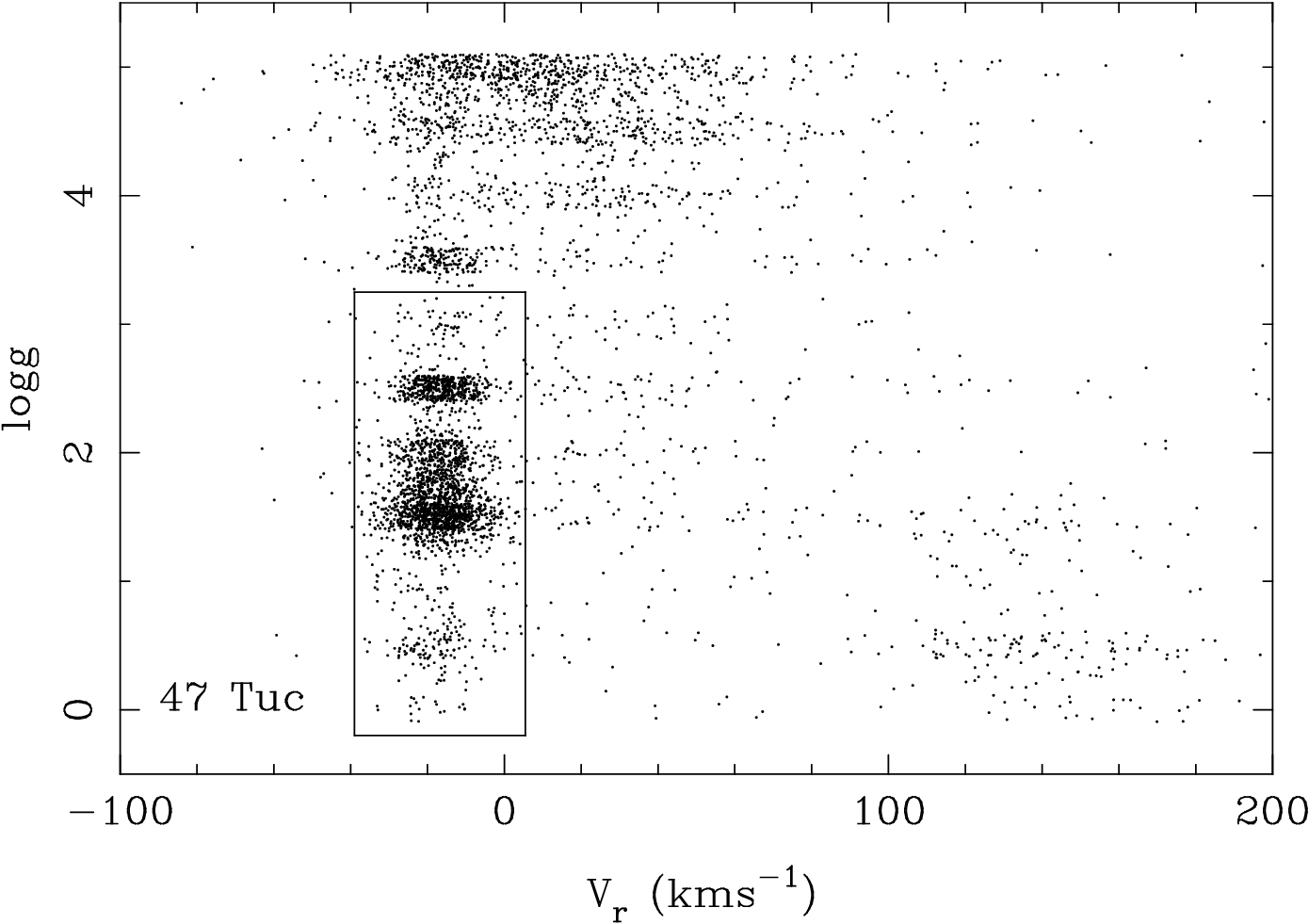}
  \includegraphics[angle=-90,width=0.45\textwidth]{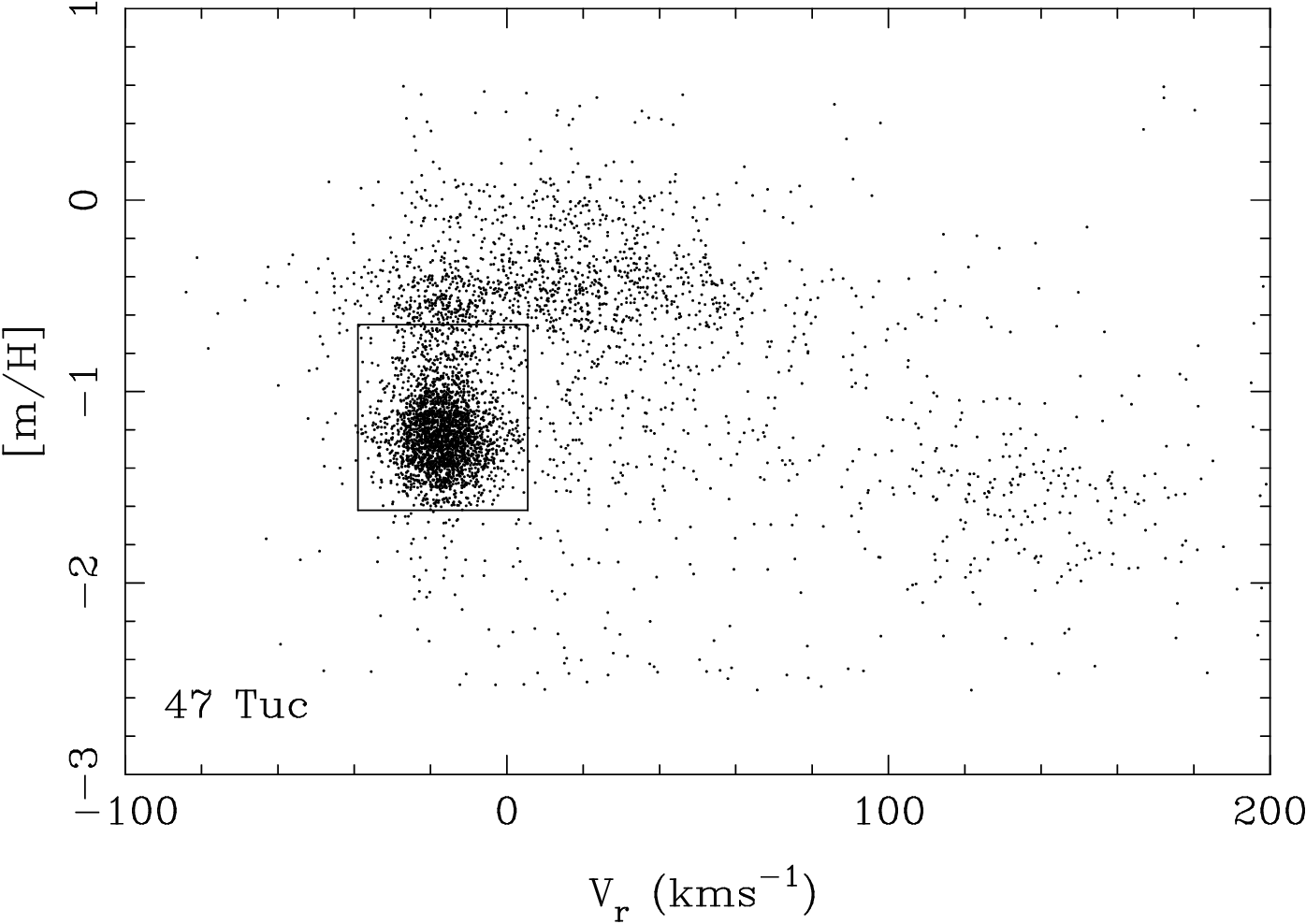}
  \includegraphics[angle=-90,width=0.45\textwidth]{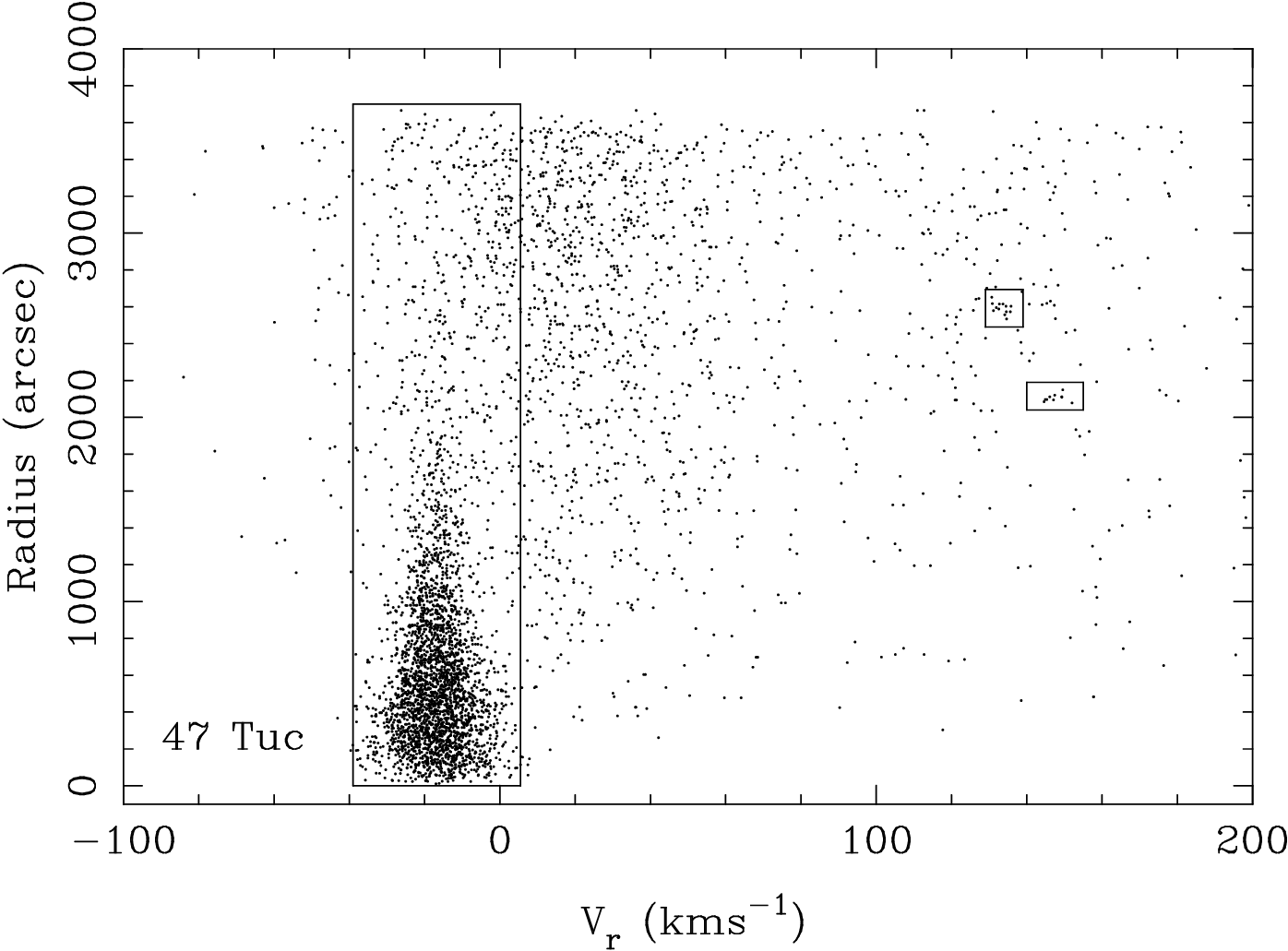}
  \includegraphics[angle=-90,width=0.45\textwidth]{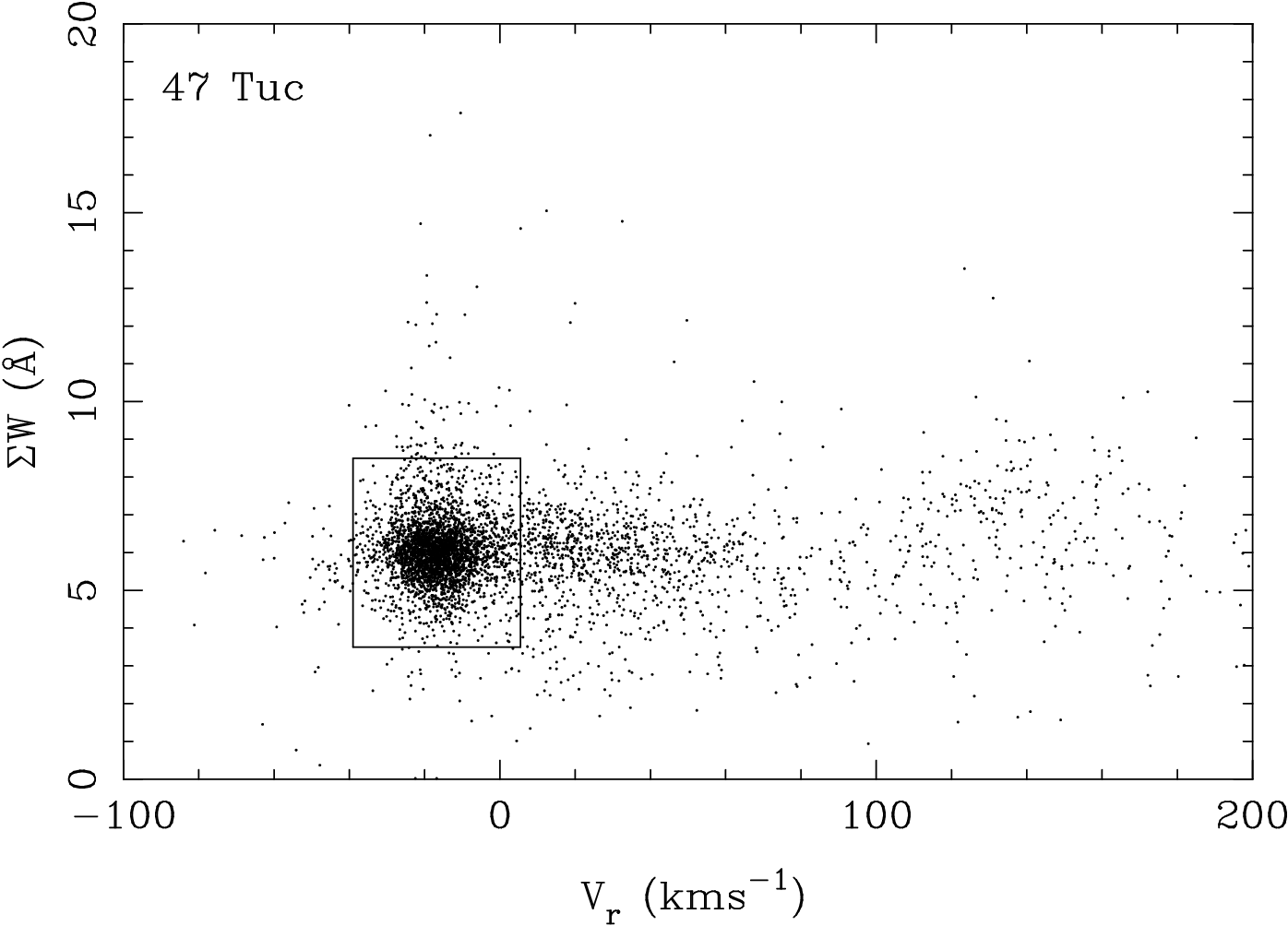}\caption{The
  selections made  for 47  Tuc. The boxes  indicate the  selections on
  each parameter. The  lower left panel also shows  the selections for
  Kron  3  (at  $\sim2600''$  radius)  and NGC  121  (at  $\sim2100''$
  radius). The selection box for Kron 3 is centered on the cluster and
  was restricted to the cluster diameter ($204''$).  For NGC 121 there
  were no stars outside $150''$  despite the diameter being similar to
  that of Kron 3.}
  \label{selections}
  \end{centering}
\end{figure*}

We  selected  cluster  members   using  four  parameters,  namely  the
equivalent width of the  calcium triplet lines, surface gravity ($\log
g$), radial velocity and metallicity ([m/H]).  Only stars that matched
all  criteria were  judged  to be  members.   A further  cut of  $\log
g<3.25$ was applied to 47 Tuc to ensure as many Galactic stars were as
possible  were  removed.   This  very probably  removed  some  cluster
members but was a necessary sacrifice to ensure our sample was as free
of Galactic  contaminants as possible.   Figure \ref{selections} shows
the selection criteria of 47 Tuc.  The selections of M55 are not shown
because the  velocity of the cluster  ($\sim175$\,km\,s$^{-1}$) is far
removed  from the Galactic  population ($\sim0$\,km\,s$^{-1}$)  so the
selections are very clean.

\begin{figure*}
  \begin{centering}
  \includegraphics[angle=-90,width=0.45\textwidth]{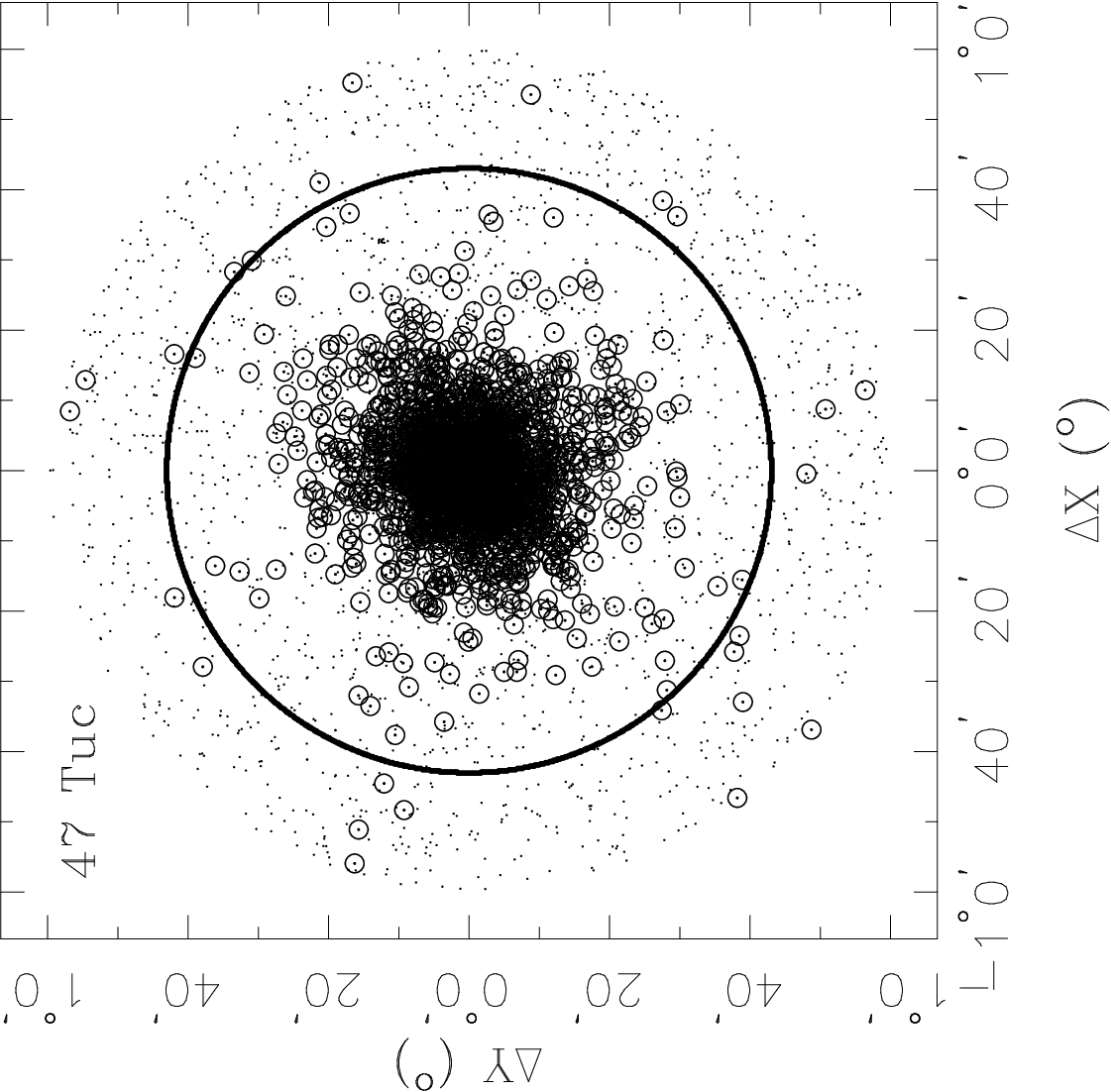}
  \includegraphics[angle=-90,width=0.45\textwidth]{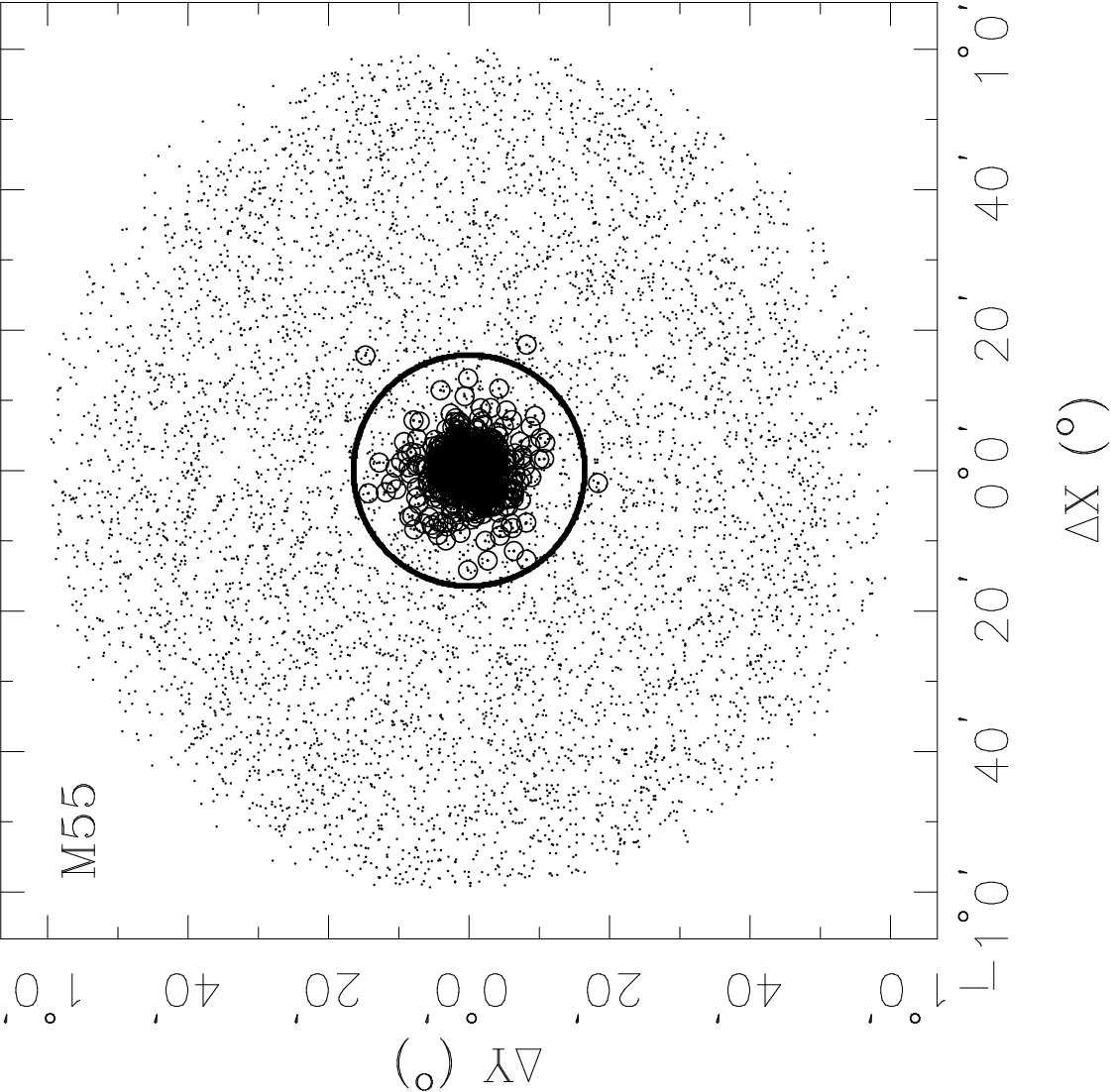}
  \caption{ The uncircled points are the stars which were observed and
    determined  not  to  be  cluster  members. The  members  used  for
    analysis, based on the selection method described in the text, are
    circled  points.  The  large circle  is  the tidal  radius of  the
    cluster from \citet{Harris96}. In each panel, North is up and East
    is to the left.}
  \label{members}
  \end{centering}
\end{figure*}

It   should  be   noted   that  for   several   clusters  studied   in
\citetalias{Lane09},  a  cutoff  of  $T_{\rm  eff}\gtrsim9000$\,K  was
necessary  to remove  hot  horizontal branch  (HB)  stars with  highly
uncertain radial  velocities because the  calcium triplet in  very hot
stars is replaced  by hydrogen Paschen lines (and  also have intrinsic
radial velocity variability,  see Section \ref{dispersions}).  No cut
was made on  $T_{\rm eff}$ for either of  the current clusters because
no  stars with  $T_{\rm eff}\gtrsim7050$\,K  (for 47  Tuc)  or $T_{\rm
eff}\gtrsim8500$\,K  (for M55) remained  after our  selection process.
Figure  \ref{members} shows  the  relative locations  of the  observed
stars and highlights those found  to be members.  Several member stars
in each cluster  were found beyond the tidal  radius; the implications
of this are discussed in Section \ref{evaporation}.

\begin{figure*}
  \begin{centering}
  \includegraphics[angle=-90,width=0.45\textwidth]{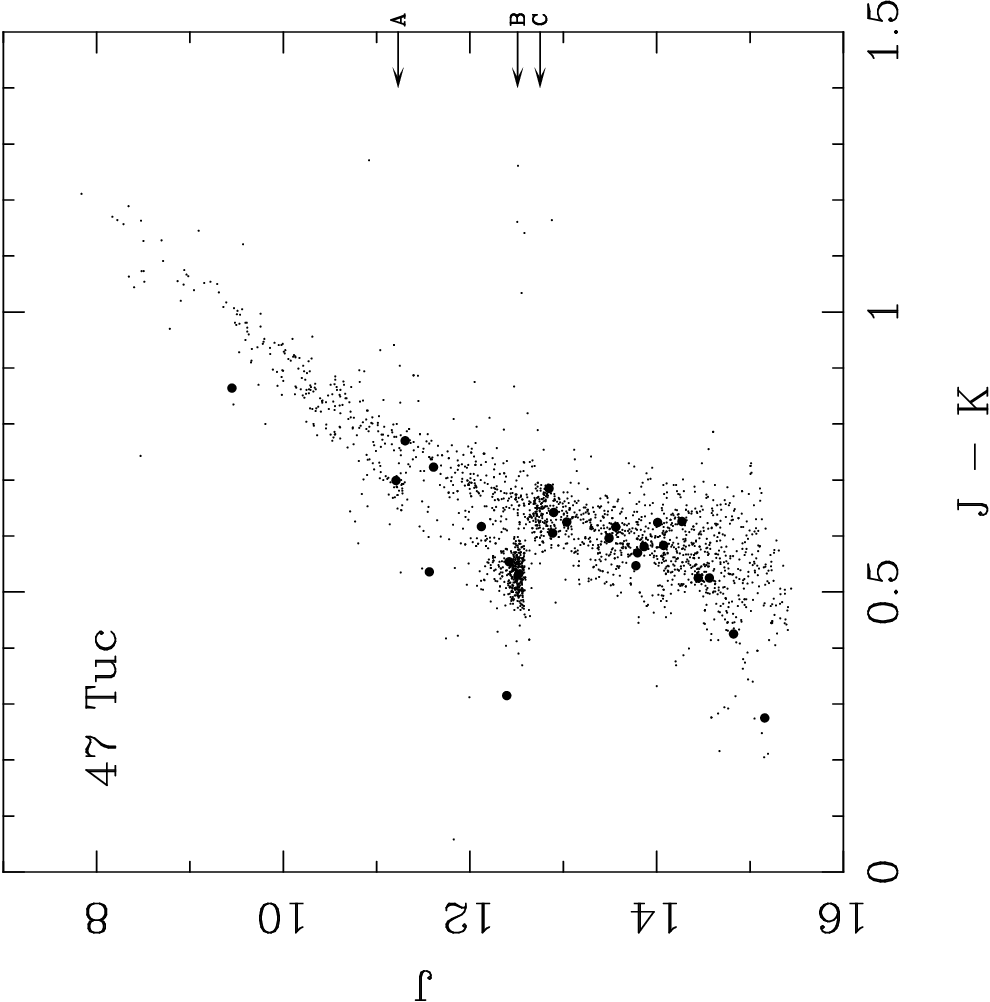}
  \includegraphics[angle=-90,width=0.45\textwidth]{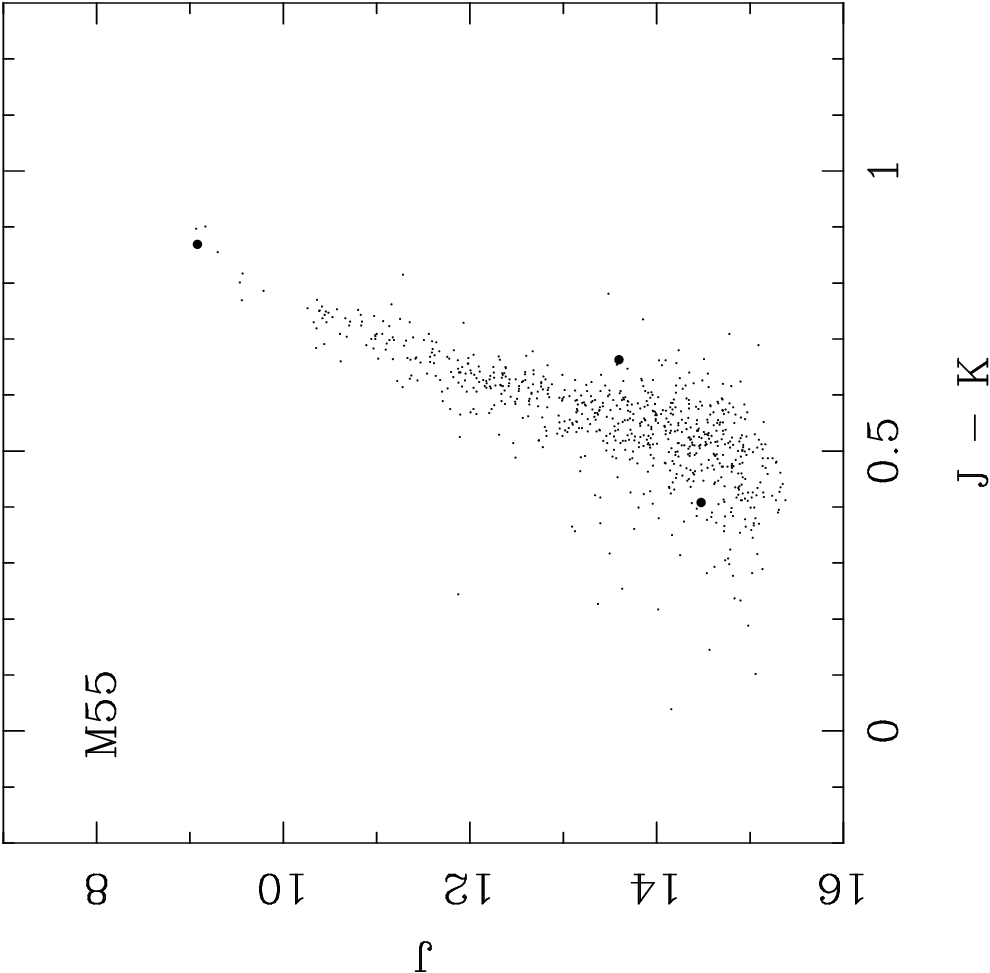}
  \caption{CMDs of selected members of 47 Tuc and M55 with extra-tidal
  stars  shown as  large points.   Because these  do not  populate any
  particular part of  the CMD, there is no  systematic contributing to
  their  selection as  members.  Note  the AGB  clump  (A), horizontal
  branch  (B)  and  RGB  bump  (C)  in  47 Tuc.   The  HB  of  M55  at
  $\sim13<J<14$ is sparsely  populated because it has a  blue HB whose
  stars are too hot for strong calcium triplet spectra.}
  \label{CMDs}
  \end{centering}
\end{figure*}

Figure \ref{CMDs}  shows the  colour-magnitude diagrams (CMDs)  of the
cluster  members, with the  extra-tidal stars  as large  points. These
extra-tidal  stars do  not populate  any specific  region of  the CMD,
indicating that there is no systematic contributing to their selection
as members.  The HB, RGB  bump and asymptotic giant branch (AGB) clump
are all  visible in 47 Tuc.   A hint of the  HB can be seen  in M55 at
$\sim13<J<14$, however,  it is not  well populated because,  unlike 47
Tuc, M55 has a blue HB, and  these stars are too hot to exhibit strong
calcium  triplet  spectra.  A  total  of  2241  and 726  members  were
selected for 47  Tuc and M55, respectively. For 47  Tuc, 98.6\% of our
final sample  used for analysis (2210  out of 2241  stars) fall within
the  99.7\%   confidence  level   for  cluster  membership   based  on
statistical analysis of each selection parameter.

\subsubsection{Sgr, NGC 121 and Kron 3}

\begin{figure}
  \begin{centering}
  \includegraphics[angle=-90,width=0.45\textwidth]{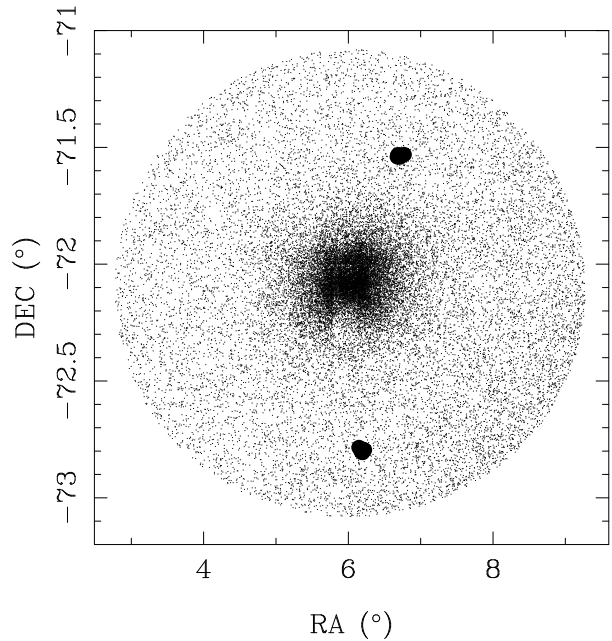}
  \caption{Selected  members of  Kron 3  (large points  below  47 Tuc:
  (RA,DEC)$\approx$($6.19$,$-72.79$)) and NGC  121 (large points above
  47 Tuc: (RA,DEC)$\approx$($6.70$,$-71.54$)).}
  \label{Kron3+NGC121members}
  \end{centering}
\end{figure}

M55 resides  in front  of the Southern  tidal tail of  the Sagittarius
dwarf galaxy \cite[Sgr;][]{Ibata94} and  42 stars from that field were
found   to  be   part  of   Sgr  (see   Section   \ref{metaldist}  for
details). These were removed  from the sample and analysed separately.
Two GCs  from the Small Magellanic  Cloud (SMC) are present  in the 47
Tuc  field,  namely NGC  121  and Kron  3.   When  determining Kron  3
membership,  we restricted  our selection  to  a box  centered on  the
cluster  with a  length of  $204''$, the  diameter of  the  cluster as
quoted  by \cite{Bica00}  (Figure \ref{selections}).   For NGC  121 we
simply selected  the clump of  stars shown in  Figure \ref{selections}
because there were no  stars outside $\sim150''$, despite the diameter
of   this  cluster  being   about  the   same  as   that  of   Kron  3
\citep{Bica00}. We then overplotted the selections on the 2MASS region
around 47  Tuc to ensure they  fell in the  same region of sky  as the
cluster    and    discarded     those    that    did    not    (Figure
\ref{Kron3+NGC121members}).  A total of 10  and 11 stars were found to
be members of NGC 121 and Kron 3, respectively.

\section[]{Results}

\subsection{Metallicity}\label{metaldist}

\begin{figure}
  \begin{centering}
  \includegraphics[angle=-90,width=0.45\textwidth]{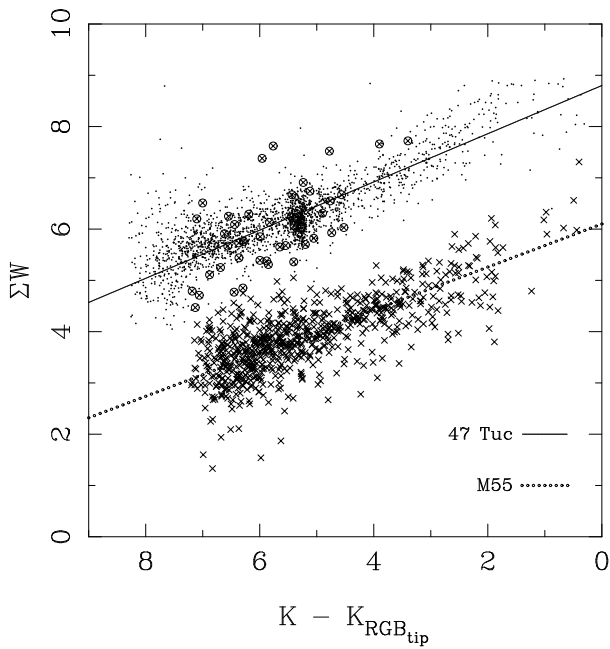}
  \caption{$K -  K_{TRGB}$ vs equivalent width of  the calcium triplet
  lines. Crosses are M55 members,  dots are 47 Tuc members and circled
  crosses are those  stars determined to belong to  the Southern tidal
  tail of Sgr. These were  analysed separately. The straight lines are
  the linear  fits to  the data  once the Sgr  members, and  all stars
  having $T_{\rm eff}>6000$\,K, were removed.}
  \label{metals1}
  \end{centering}
\end{figure}

Given the large sample sizes of  both clusters we were able to perform
an  accurate metallicity  calibration  in  a similar  way  to that  by
\cite{Cole04}  and  \cite{Warren09}.  Our  method  uses  the TRGB  $K$
magnitude ($K_{TRGB}$),  instead of the  HB used by  \cite{Cole04} and
\cite{Warren09}.  Both  methods are robust,  but the TRGB  is brighter
than the HB, allowing our method  to be used for more distant clusters
where the HB is  not visible. In addition, only in the  $K$ band has a
direct  calibration of  the TRG  been made  with  Hipparcos parallaxes
\citep{Tabur09}.  The metallicity calibration was carried out in three
steps.  Firstly, $K_{TRGB}$ was  subtracted from each star and plotted
against the equivalent width of  the calcium triplet lines (see Figure
\ref{metals1}),   giving  a  distance   independent  measure   of  the
luminosity.   $K_{TRGB}$ values were  taken from  \cite{Valenti04} (47
Tuc,  M30 M55  and M68),  \cite{Marconi98} (Sgr  core) and  2MASS CMDs
(Kron 3, NGC 121, M22 and M53; see below).  Several stars from the M55
field  overlapped the  47  Tuc  region in  this  figure.  Because  the
metallicity of  47 Tuc is  similar to that  of the tidal tails  of Sgr
\cite[e.g.][]{Chou07}, this  overlap was expected;  these stars belong
to  the Southern  tidal  tail of  Sgr  and are  highlighted in  Figure
\ref{metals1}.  All stars with $T_{\rm eff}>6000$\,K were removed from
the M55  sample for the  metallicity analysis because these  hotter HB
stars  have the  calcium triplet  lines affected  by  hydrogen Paschen
lines  and, therefore, should  not be  used for  metallicity analysis.
This was not necessary for 47 Tuc since the selections were restricted
to stars  with $\log  g<3.25$, because this  cluster is closer  to the
Galaxy in $\log g$, which removed all the hotter HB stars.

The second step  in the calibration process was  to fit straight lines
to  the  data  (Figure  \ref{metals1}).   Because our  47  Tuc  sample
contains many  HB stars,  and these were  not expected to  exhibit the
same relation between calcium triplet  line widths and $K - K_{TRGB}$,
these fits  were also performed with  the HB stars of  47 Tuc removed.
No difference in  the fits was found, so the HB  stars are included in
Figure \ref{metals1}.  The slope of  these lines are 0.47 and 0.42 for
47 Tuc and M55, respectively ($cf.$  0.64 for $V$ band analysis of the
HB  by \citealt{Cole04}  and  0.47 for  $K$  band analysis  of the  HB
\citealt{Warren09}.).

\begin{figure}
  \begin{centering}
  \includegraphics[angle=-90,width=0.45\textwidth]{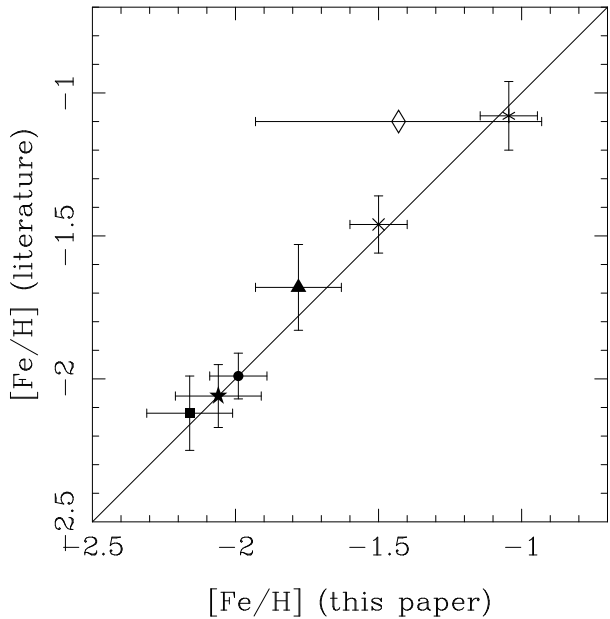}
  \caption{Calculated  [Fe/H]  for   M30  (square),  M68  (star),  M53
  (circle),  M22 (triangle), NGC  121 (cross),  Sgr (diamond),  Kron 3
  (asterisk)   versus  literature  values   (see  text).    The  large
  uncertainty for Sgr is due to the difficulty in calculating the TRGB
  for this  object. No consensus  on an uncertainty estimate  has been
  reached, hence none is shown here from the literature.}
  \label{metals4}
  \end{centering}
\end{figure}

Thirdly,    by     plotting    [Fe/H]    of     the    two    clusters
\cite[from][]{Harris96} vs $\Sigma W -  AX$, where $A$ is the gradient
of the slope  above and $X$ is $K - K_{TRGB}$,  for these two clusters
we have a calibrator on [Fe/H] \cite[see][for a detailed discussion of
this calibration  methodology]{Cole04,Warren09}.  $\Sigma W  - AX$ can
then  be calculated  for  any cluster  and  therefore [Fe/H].   Figure
\ref{metals4}  shows [Fe/H]  calculated by  this method  versus [Fe/H]
from   the    literature:   for   Sgr    \cite[][]{Chou07},   Kron   3
\cite[][]{Glatt08b}, NGC  121 \cite[][]{Glatt08a} as well  as the four
clusters from \citetalias{Lane09}, namely M22 \cite[][]{Monaco04}, M30
\cite[][]{Harris96},      M53     \cite[][]{Harris96}      and     M68
\cite[][]{Harris96} (TRGB  values for  these final four  clusters were
all measured from 2MASS CMDs),  showing this calibration for [Fe/H] to
be robust.   For clarity, the [Fe/H]  values from this  paper, and the
literature, are also shown in Table \ref{metaltable}.

\begin{table}
\begin{center}
\caption{[Fe/H]  values from this  paper and  from the  literature, in
  order  of decreasing  metallicity. The  literature values  are taken
  from:   Kron  3  \citep{Glatt08b},   Sgr  \citep{Chou07},   NGC  121
  \citep{Glatt08a},  M22 \citep{Monaco04},  M53  \citep{Harris96}, M68
  \citep{Harris96} and M30 \citep{Harris96}.}\label{metaltable}
\begin{tabular}{@{}ccc@{}}
\hline
\hline
Cluster & [Fe/H] (this paper) & [Fe/H] (literature)\\
\hline
Kron 3 & $-1.05\pm0.1$ & $-1.08\pm0.12$\\
Sgr & $-1.4\pm0.5$ & $\sim-1.1$\\
NGC 121 & $-1.5\pm0.1$ & $-1.46\pm0.1$\\
M22 & $-1.78\pm0.15$ & $-1.68\pm0.15$\\
M53 & $-1.99\pm0.1$ & $-1.99\pm0.08$\\
M68 & $-2.06\pm0.15$ & $-2.06\pm0.11$\\
M30 & $-2.16\pm0.15$ & $-2.13\pm0.13$\\
\hline
\end{tabular}
\end{center}
\end{table}

\begin{figure}
  \begin{centering}
  \includegraphics[angle=-90,width=0.45\textwidth]{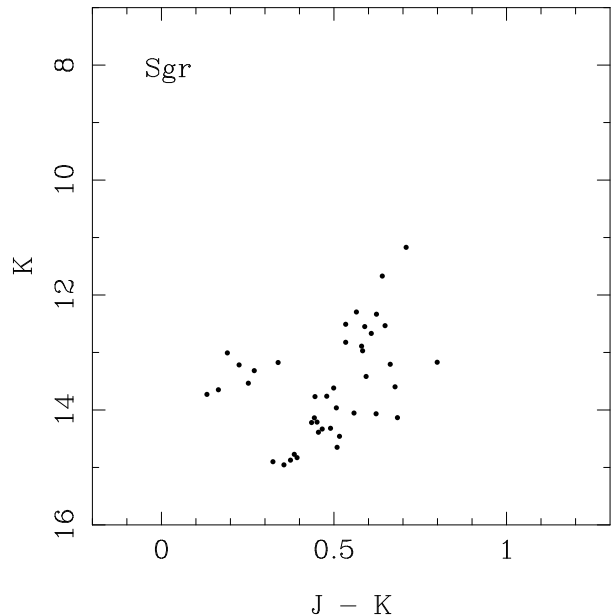}
  \caption{CMD of Sgr  stars extracted from the M55  field showing the
  TRGB at $K\approx11.15$.}
  \label{SgrCMDs}
  \end{centering}
\end{figure}

The large uncertainty for Sgr  is due to the difficulty in determining
an accurate measure of the TRGB for this object, because the Sgr tidal
tails   are   close   to   the  Galaxy   in   colour-magnitude   space
\cite[e.g.][]{Marconi98}. Furthermore, there is a metallicity gradient
along the tails.   To determine the $K$ magnitude of  the TRGB for the
Southern Sgr  tail at the location  covered in the  current survey, we
produced  a  CMD  from  our  data  (Figure  \ref{SgrCMDs})  and  found
$K_{TRGB}\approx11.15$.  The  paucity of stars in  this CMD introduces
large uncertainties into the calculated [Fe/H] for this object (Figure
\ref{metals4}).

Since no literature values of the $K_{TRGB}$ are available for Kron 3,
NGC 121, M22 or M53, we  produced CMDs from 2MASS of stars within $2'$
of the cluster core (for M22 and M53) and within $10''$ for Kron 3 and
NGC  121. Several  hundred stars  in  the CMDs  for M22  and M53,  and
$\sim100$ for Kron 3 and NGC  121, meant that accurate values could be
calculated.


\subsection{Rotation}\label{rotation}

\begin{figure}
  \begin{centering}
  \includegraphics[angle=-90,width=0.47\textwidth]{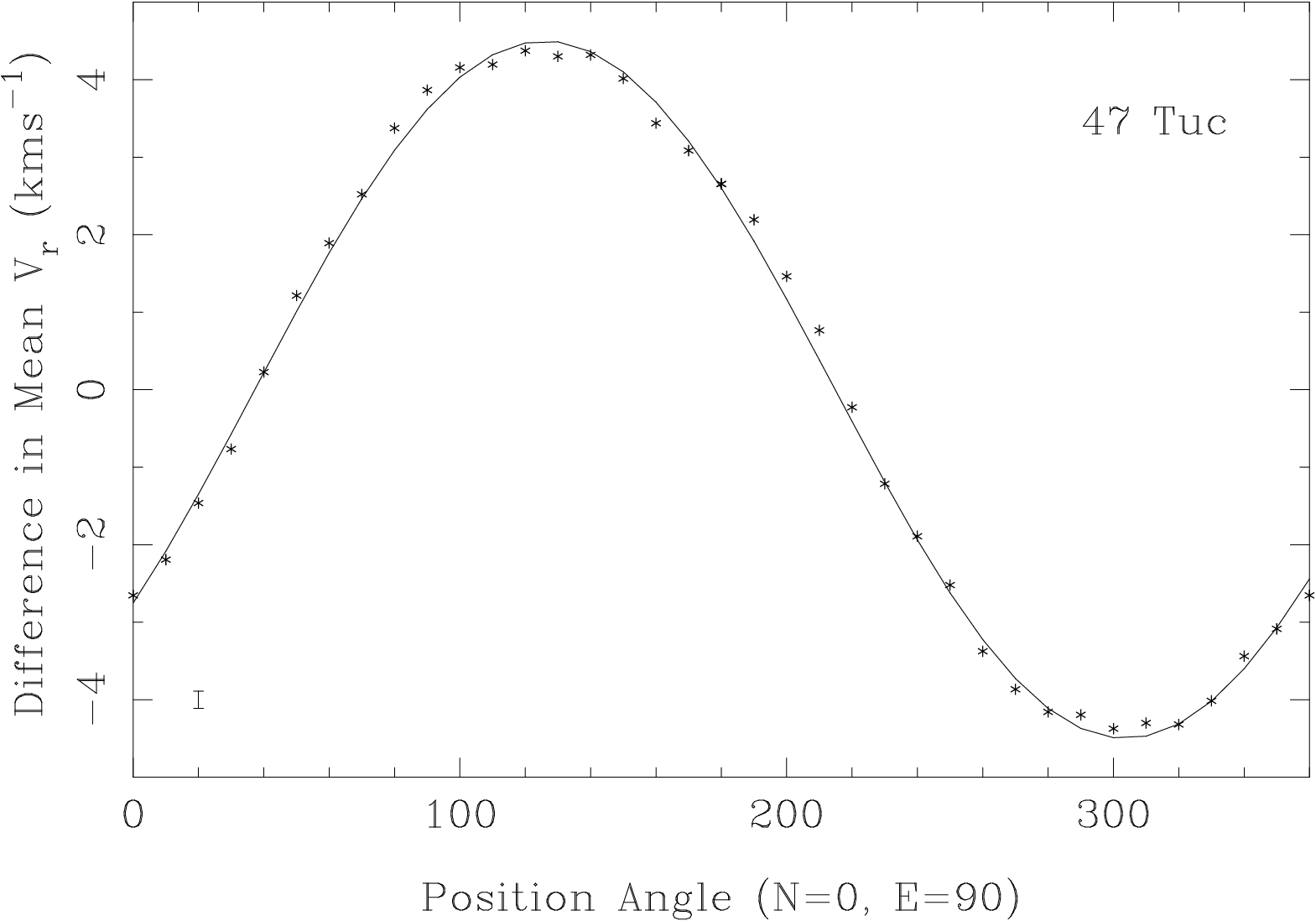}
  \includegraphics[angle=-90,width=0.47\textwidth]{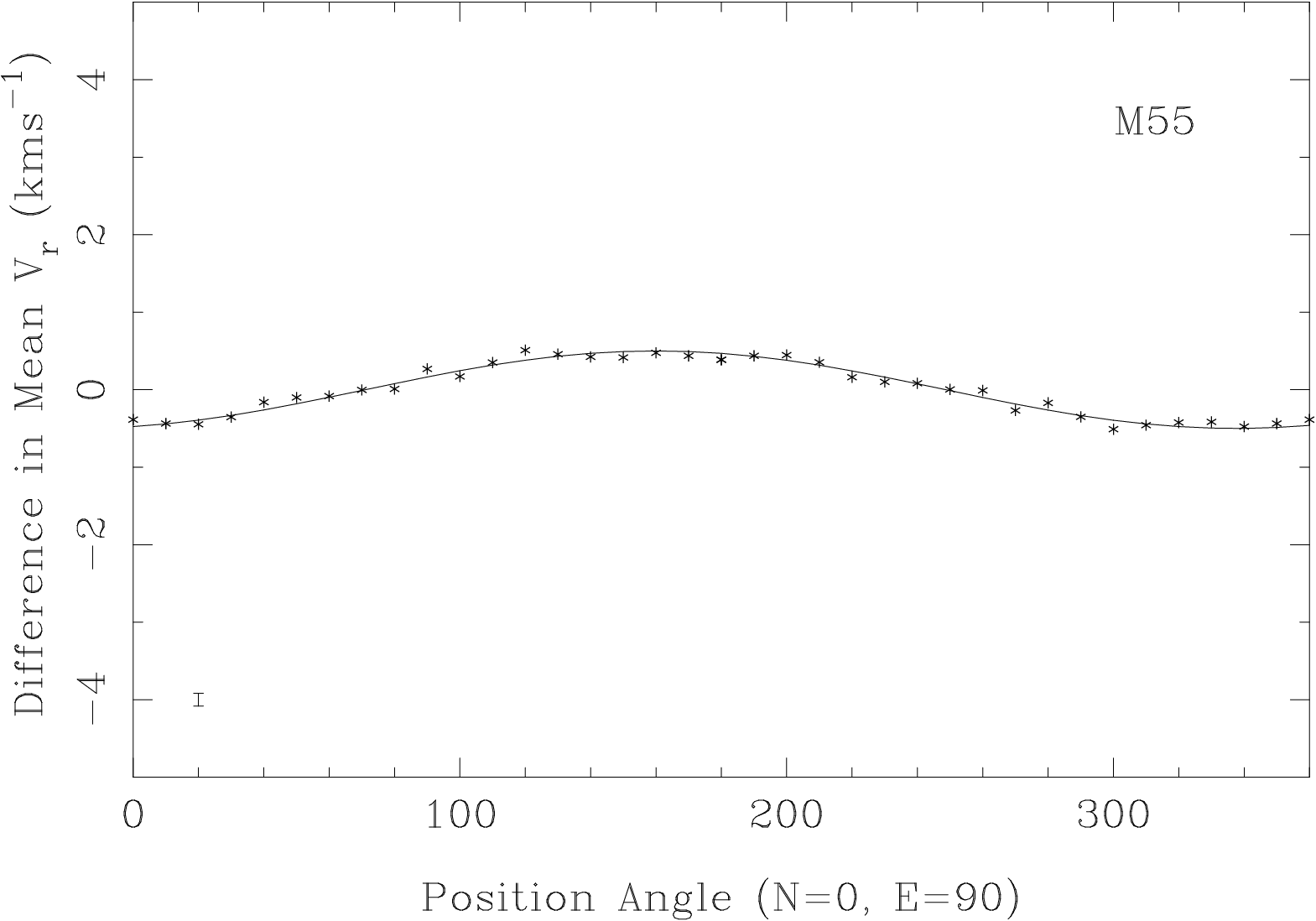}
  \caption{The rotation  of each cluster calculated  as the difference
    between  the mean  velocities on  each side  of the  cluster along
    equal  position  angles,  as  described  in the  text.   The  best
    $\chi^2$ fit sine function is overplotted, and a typical error bar
    is represented in the lower left of each panel.}
  \label{rotate}
  \end{centering}
\end{figure}

Assuming each cluster has  an isothermal distribution, their rotations
were  measured by  halving  the  cluster by  position  angle (PA)  and
calculating  the mean  radial velocity  of  each half.   The two  mean
velocities  were then  subtracted.   This was  performed  in steps  of
$10^\circ$,  and  the  best-fitting  sine function  overplotted.   The
results are presented in Figure \ref{rotate}.

This  method gives  an amplitude  that  is twice  the actual  measured
rotation.      Therefore,    47     Tuc    exhibits     rotation    at
$2.2\pm0.2$\,km\,s$^{-1}$  with  an  approximate projected  rotational
axis along the line PA  = $40^\circ-220^\circ$, and M55 shows rotation
at a  level of $0.25\pm0.09$\,km\,s$^{-1}$ and an  approximate axis of
rotation  along  the line  PA  =  $65^\circ-245^\circ$.  Our  rotation
measure   for  47  Tuc   corresponds  to   the  value   calculated  by
\cite{Meylan86}    at   a   radius    of   $\approx20$\,pc    and   by
\cite{Strugatskaya88}  at  $\approx36$\,pc.   \cite{Szekely07}  showed
that M55 is rotating with a velocity of $\sim0.5$\,km\,s$^{-1}$.  This
discrepancy can  probably be  attributed to \cite{Szekely07}  having a
sample size  approximately half that  of the current study.   For both
clusters, we  corrected the individual  stellar velocity data  for the
measured rotation  before the velocity dispersions,  and M/L$_{\rm V}$
profiles, were calculated.

\subsection{Velocity Dispersions}\label{dispersions}

\begin{figure}
  \begin{centering}
  \includegraphics[angle=-90,width=0.47\textwidth]{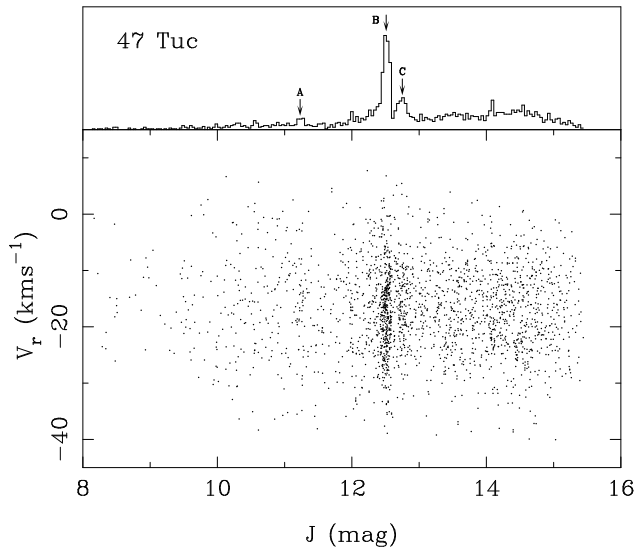}
  \caption{Velocity versus $J$ magnitude of the members of 47 Tuc. The
  HB stars (B)  appear to have the largest  velocity dispersion of any
  stellar type.  As  per Figure \ref{CMDs}, the AGB  clump is labelled
  'A', the horizontal branch 'B' and the RGB bump 'C'.}
  \label{47TucvelvsJ}
  \end{centering}
\end{figure}

\begin{figure}
  \begin{centering}
  \includegraphics[angle=-90,width=0.45\textwidth]{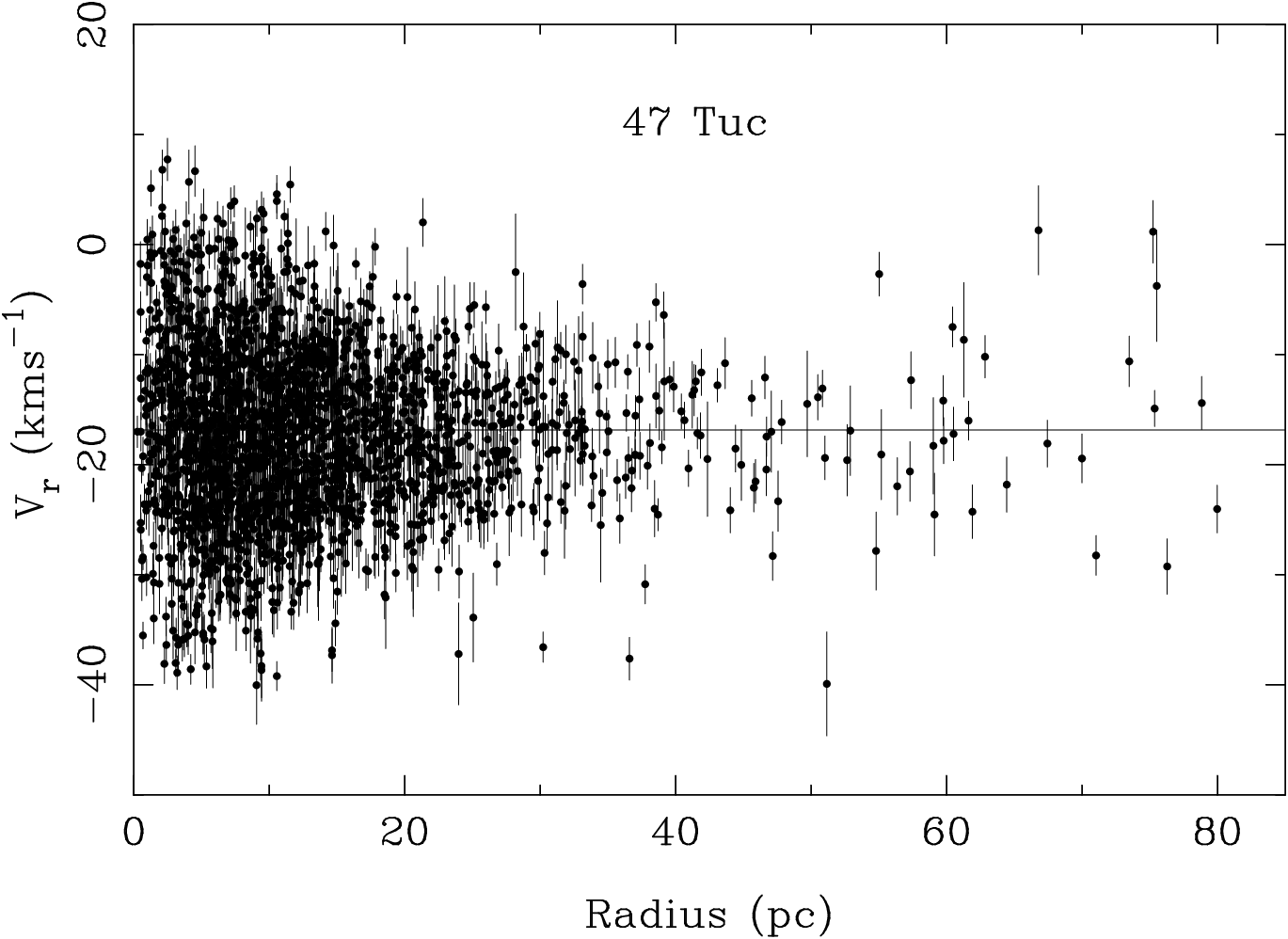}
  \includegraphics[angle=-90,width=0.45\textwidth]{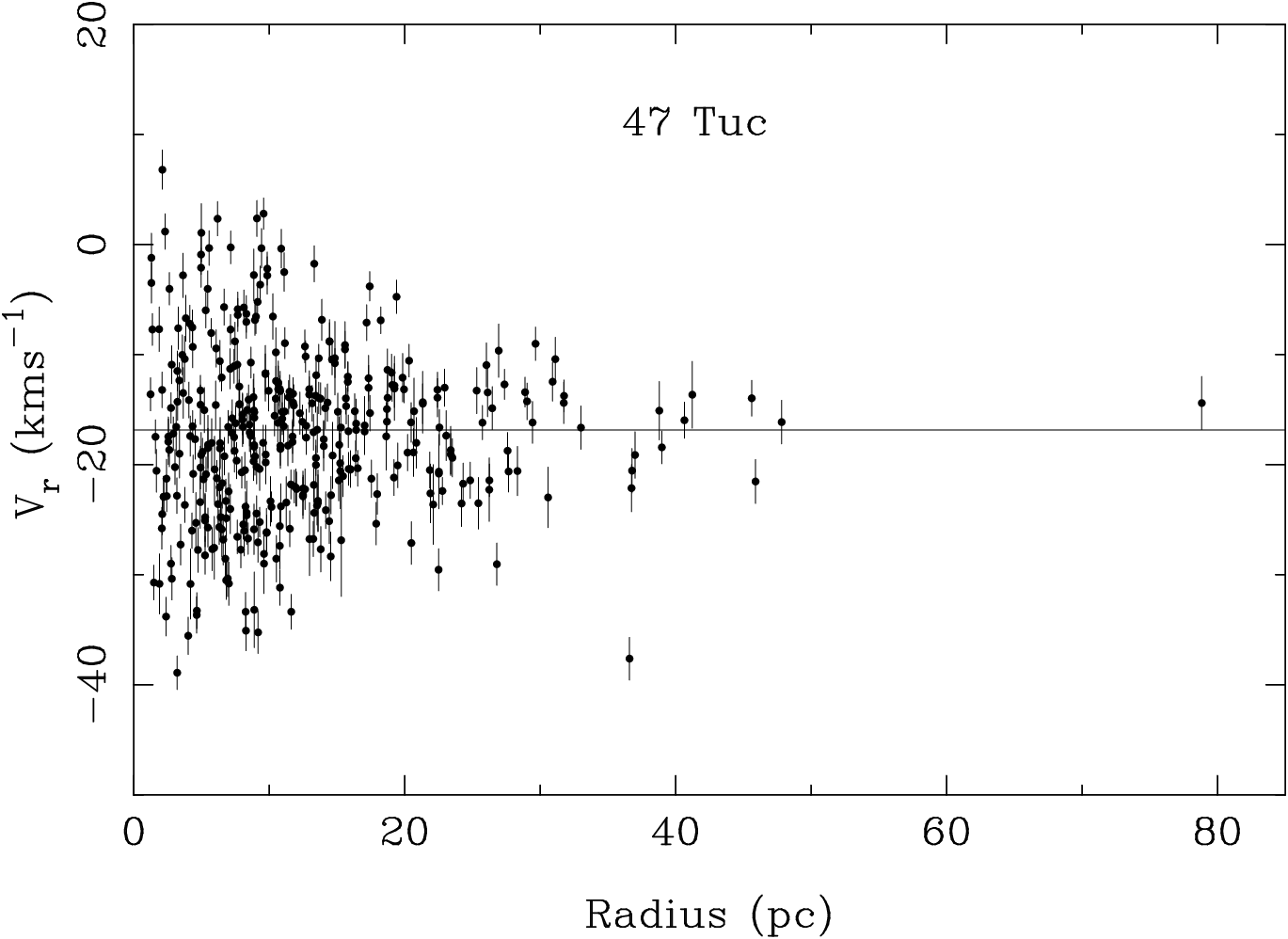}
  \caption{Velocities,  and associated uncertainties,  versus distance
  of 47  Tuc. Complete  sample (top panel)  and HB stars  only (bottom
  panel). Note that the HB stars are evenly distributed throughout the
  cluster.   The   horizontal  line  denotes   the  measured  systemic
  velocity.}
  \label{47Tucvelvsdist}
  \end{centering}
\end{figure}

Figure \ref{47TucvelvsJ}  shows velocity versus $J$  magnitude for the
47 Tuc data. It appears that  the HB population (labelled 'B') has the
greatest velocity dispersion of  any stellar population in our sample.
Stellar pulsations can alter the  velocity dispersion profile of a GC,
if  pulsating  stars are  present  in  sufficient  numbers, so  it  is
important to check for this effect.   47 Tuc only has one single known
RR Lyrae star \cite[][]{Bono08} so this will not affect the dispersion
profile, however, it is natural to  expect that the large number of HB
stars in  this sample to have an  effect, since HB stars  are known to
pulsate when located close to  the instability strip.  To test this we
calculated  the  velocity dispersion  of  the  47  Tuc sample  in  $J$
magnitude bins.  The bin containing the HB stars ($12.43<J<12.59$) has
a  velocity dispersion of  $7.5\pm0.3$ and  the overall  dispersion is
$7.5\pm0.6$.   Because  these  stars   do  not  show  an  increase  in
dispersion compared  with the  complete sample, and,  furthermore, are
distributed     evenly      throughout     the     cluster     (Figure
\ref{47Tucvelvsdist}), we see  no reason to exclude the  HB stars from
the velocity dispersion analysis.  Our M55 sample contains very few HB
stars (see  Figure \ref{CMDs})  so this effect  is negligible  in this
cluster.

We  measured the  systemic velocity  of each  cluster, using  a Markov
Chain Monte Carlo  (MCMC) maximum likelihood method \citep{Gregory05},
which takes into account  the individual velocity uncertainties on the
stars, providing the  systemic velocity with associated uncertainties.
A comparison between our values and and those from the \cite{Harris96}
catalogue  are   shown  in  Table   \ref{velocity  comparisons}.   The
velocities   show    systematic   differences   of    the   order   of
$2-3$\,km\,s$^{-1}$   between   our   mean   values   and   those   in
\cite{Harris96}.  These differences are similar to what \cite{Balog09}
found  for  two  open  clusters,  NGC  2451A and  B,  using  the  same
instrument and analysis method and comparing data to velocities in the
literature.  Our  interpretation is that  there might be  a systematic
uncertainty in  the zero-point of our velocity  system.

\begin{table}
\begin{center}
\caption{Comparisons  between the systemic  radial velocities  of each
  cluster  in  the  \citet{Harris96}  catalogue and  those  from  this
  survey. Velocities are in km\,s$^{-1}$.}\label{velocity comparisons}
\begin{tabular}{@{}cccc@{}}
\hline
\hline
Cluster &  V$_r$ \citep{Harris96} & V$_r$ (this paper)\\
\hline
47 Tuc & -18.7$\pm0.2$ & $-16.85\pm0.16$\\
M55 & 174.8$\pm0.4$ & $177.37\pm0.13$\\
\hline
\end{tabular}
\end{center}
\end{table}

To determine  the velocity dispersions  of our samples, we  binned the
stars in annuli centered on the cluster centre, ensuring approximately
equal numbers of stars per bin  ($\sim50$ for M55 and $\sim100$ for 47
Tuc; see  Table \ref{bins}  for the bin  dimensions). The  MCMC method
described above was then used to determine the dispersion in each bin,
and the  resulting velocity dispersion profiles  were overplotted with
the  best-fitting   \cite{Plummer11}  model.   The   central  velocity
dispersion and  the scale radius  ($r_s$; containing half  the cluster
mass) were  used for  the fitting.  The  Plummer model allows  for the
calculation  of  the total  cluster  mass  from  the central  velocity
dispersion ($\sigma_0$) and $r_s$  via (see \citealt{Dejonghe87} for a
discussion of Plummer models and their application):

\[
M_{tot} = \frac{64\sigma_0^2r_s}{3\pi G}.
\]

\begin{table}
\begin{center}
\caption{Dimensions  of the  bins\new{,  and the  number  of stars  in
each}, used in  the velocity dispersion analysis.  Only  the inner bin
edges  are shown.  Values given  are parsecs  from the  centre  of the
cluster. The  final bin extends  to 80\,pc for  47 Tuc and  35\,pc for
M55. }\label{bins}
\begin{tabular}{@{}cccc@{}}
\hline
\hline
\multicolumn{2}{c}{47 Tuc} & \multicolumn{2}{c}{M55}\\
Inner bin edge & Stars per bin & Inner bin edge & Stars per bin\\
\hline
 0.000 & 99 & 0.00 & 51\\
 2.131 & 102 & 1.41 & 45\\
 3.048 & 99 & 1.98 & 49\\
 4.031 & 99 & 2.62 & 52\\
 4.834 & 101 & 3.14 & 49\\
 5.698 & 100 & 3.76 & 51\\
 6.416 & 100 & 4.29 & 45\\
 7.210 & 100 & 4.71 & 55\\
 8.080 & 100 & 5.32 & 47\\
 8.894 & 100 & 6.05 & 50\\
 9.699 & 100 & 6.89 & 49\\
 10.600 & 100 & 7.73 & 52\\
 11.460 & 100 & 8.97 & 49\\
 12.409 & 100 & 11.12 & 44\\
 13.347 & 99 & 15.00 & 38\\
 14.713 & 101 & & \\
 15.849 & 100 & & \\
 17.647 & 100 & & \\
 19.484 & 100 & & \\
 21.899 & 100 & & \\
 25.069 & 100 & & \\
 30.220 & 100 & & \\
 46.704 & 41 & & \\
\hline
\end{tabular}
\end{center}
\end{table}

\begin{figure}
  \begin{centering}
  \includegraphics[angle=-90,width=0.47\textwidth]{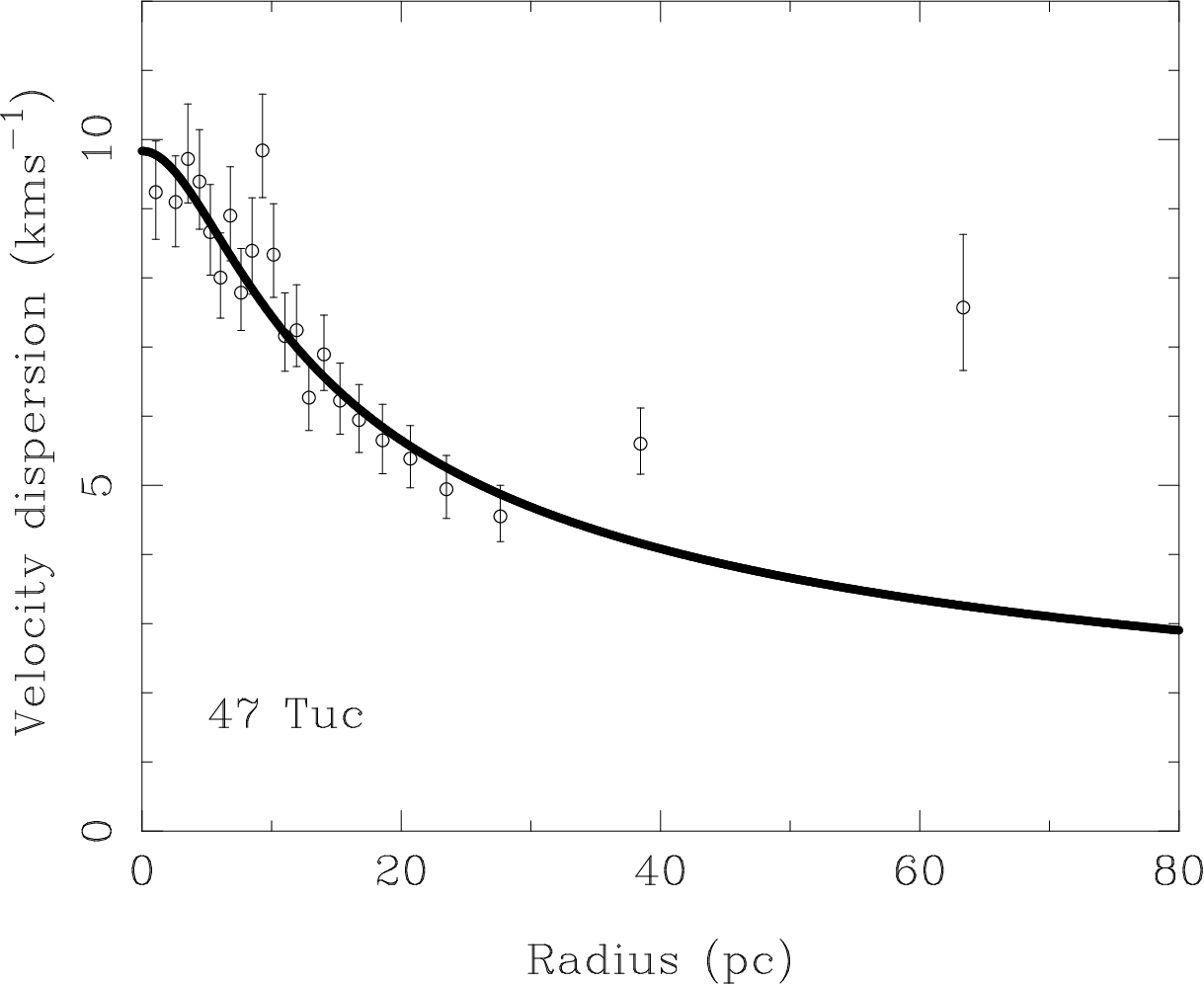}
  \includegraphics[angle=-90,width=0.47\textwidth]{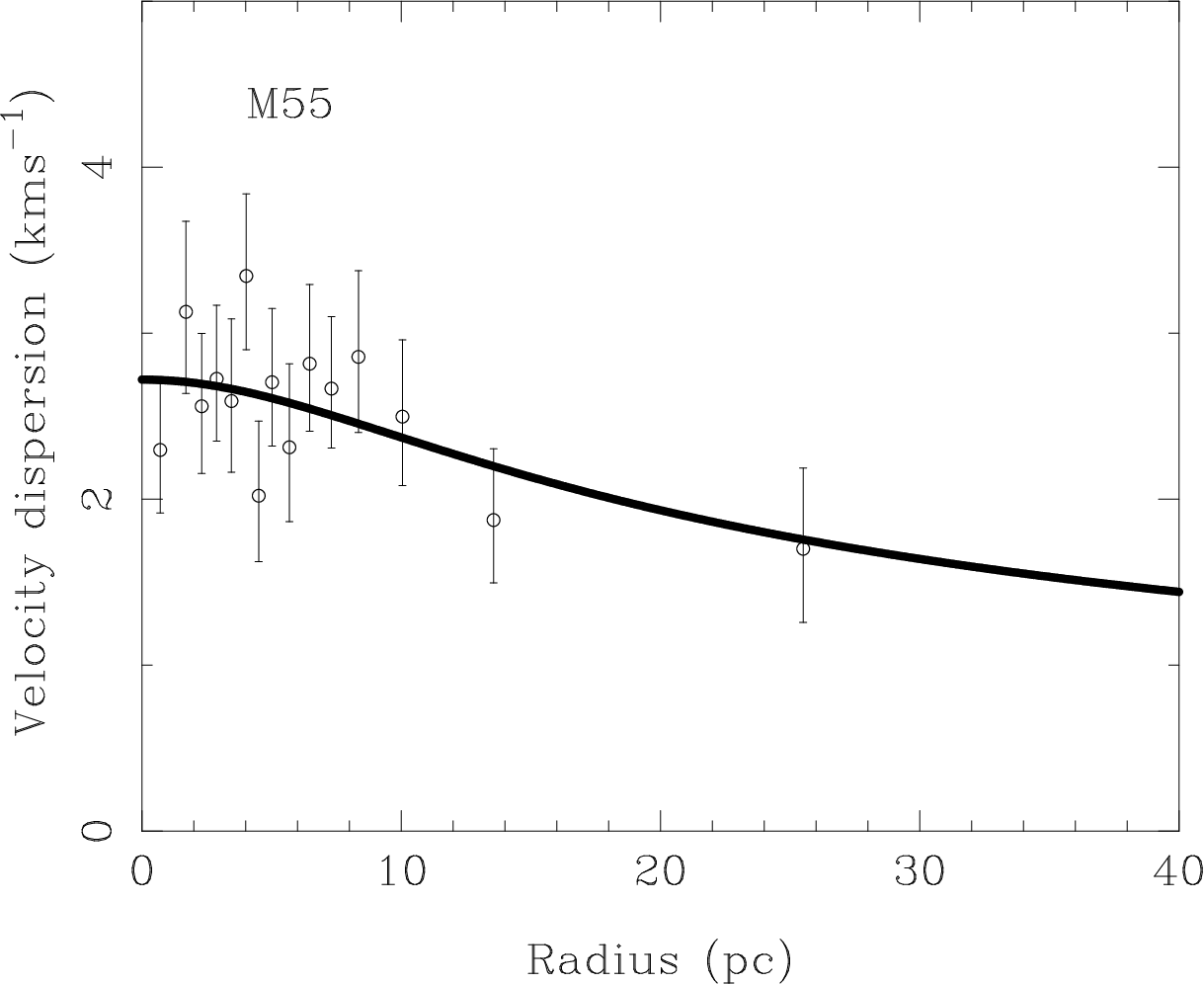}
  \caption{Velocity  dispersion  profiles of  each  cluster. The  best
  fitting \citet{Plummer11} model is overplotted. Note that for 47 Tuc
  the outermost 2 bins are  not included during the fitting process to
  ensure no artificial increase in  the estimates of total mass, $r_s$
  or central dispersion.}
  \label{veldisp}
  \end{centering}
\end{figure}

Twenty  five extra-tidal  stars were  found to  be members  of  47 Tuc
($r_t\sim56$\,pc; \citealt{Harris96}).  The  velocity dispersion of 47
Tuc  shows  a  marked  increase  for  $R\gtrsim28$\,pc  (\new{this  is
discussed in}  Section \ref{evaporation}). Because of  this increase in
velocity  dispersion in  the outer  regions, the  outer two  bins were
excluded  from the Plummer  model fitting  to the  dispersion profiles
(Figure  \ref{veldisp}), since including  them in  the fit  would have
created an artificial increase in the fit over the entire cluster, and
hence artificially  altered the total  mass, scale radius  and central
dispersion estimates.  For M55, only 3 extra-tidal stars were found to
be members, and since there  is no increase in the dispersion profile,
these are not  affecting the profile in the  outskirts of the cluster.
The  total masses, scale  radii and  central velocity  dispersions are
presented in  Table \ref{dispparams}.   Our mass estimates  agree well
with                           other                           studies
\cite[e.g.][]{Meylan89,Pryor93,Meziane96,Kruijssen09},  none  of  whom
used Plummer models to calculate their estimates.

\begin{table}
\begin{center}
\caption{Total  masses, scale radii ($r_s$)  and central  velocity
  dispersions ($\sigma_0$)
  for 47 Tuc and M55.}\label{dispparams}
\begin{tabular}{@{}cccc@{}}
\hline
\hline
Cluster & Total mass ($M_\odot$) & $r_s$ (pc) & $\sigma_0$ (km\,s\,$^{-1}$)\\
\hline
47 Tuc & $1.1\pm0.1\times10^6$ & $7.8\pm0.9$ & $9.6\pm0.6$\\
M55 & $1.4\pm0.5\times10^5$ & $11.7\pm4.2$ & $2.7\pm0.5$\\
\hline
\end{tabular}
\end{center}
\end{table}

\citet{Scarpa03,Scarpa04a,Scarpa04b} showed  an apparent flattening of
the  velocity dispersion  profiles of  five  of the  six GCs  studied,
indicating  a  significant  DM  component,  or a  modified  theory  of
gravity, was required to  explain their results.  MOG models generally
differ from Newtonian gravity for large accelerations (e.g.  in galaxy
clusters or elliptical galaxies) but become Newtonian for intermediate
accelerations [e.g.   for solar system bodies;  see \cite{Moffat08} as
an example].   MOND, however, becomes  non-Newtonian for accelerations
below    about   $1.2\times10^{-10}$\,m\,s$^{-2}$   \citep{Milgrom83},
approximately the regime where dark  matter is invoked to explain, for
example, rotation curves of galaxies.

Plummer models  have the  advantage of being  monotonically decreasing
and, therefore,  any flattening of the profiles  would be discernible.
Within the  limits of  the model, and  the uncertainties in  the data,
neither  velocity  dispersion exhibits  any  flattening, although  the
profile of 47  Tuc exhibits an increase in  its dispersion. This could
not  be called  ``flattening'' and  a MOND/MOG  model, or  DM,  is not
required to explain this  phenomenon\new{, although a DM component may
be  one explanation}  (see Section  \ref{evaporation}).  Therefore, we
infer  that neither  dark matter,  nor a  modification to  the current
understanding of  gravitation, are needed  to explain the  dynamics of
either   47   Tuc   or   M55,   corroborating   earlier   results   in
\citetalias{Lane09} for M22, M30, M53 and M68, and similar conclusions
drawn by \cite{Sollima09} for $\omega$\,Centauri.

Because  we only  sampled 10  and 11  stars from  NGC 121  and  Kron 3
respectively,  it was  not possible  to create  a  velocity dispersion
$profile$  for these  objects. Instead,  a single  velocity dispersion
value for the  cluster was calculated, with NGC  121 having a velocity
dispersion    of    $2.2\pm1.1$\,km\,s$^{-1}$    and   Kron    3    of
$1.8\pm0.9$\,km\,s$^{-1}$.

\subsubsection{Evaporation}\label{evaporation}

GCs are  known to  be tidally destroyed  by their host  galaxy through
tidal  heating   \cite[e.g.   Pal  5;][]{Odenkirchen03}.    The  tidal
stripping  signature  is evidenced  by  stars  being  stripped in  two
directions  (the  leading and  trailing  tidal  streams). Because  the
sampled  field around  47 Tuc  does not  reach far  outside  the tidal
radius (Figure \ref{members}), it is  not possible to tell whether the
extra-tidal  stars exhibit  a preferential  direction.   Therefore, an
inspection of a $20^\circ\times20^\circ$ region centered on 47 Tuc was
performed,   using   2MASS  data   selected   on   the   RGB  and   HB
colours/magnitudes  of  47  Tuc.    No  evidence  for  extended  tidal
structure  was  observed,   independently  confirming  the  result  by
\cite{Leon00} who  found no  convincing statistical evidence  of tidal
tails.

Evaporation in  GCs has been shown  to occur due  to internal two-body
relaxation  over  the  cluster  lifetime  \cite[e.g.][]{McLaughlin08},
particularly during core-collapse, or in post-core-collapsed clusters.
Importantly, $N$-body simulations have shown this evaporation exhibits
a  signature  in  the  velocity  dispersion  profile,  increasing  the
dispersion at $r_t/2$,  precisely the region where our  47 Tuc profile
increases  (\citealt{Drukier07};  for  47 Tuc,  $r_t/2\approx28$\,pc).
Furthermore, the extra-tidal  velocity distribution is symmetric about
the  systemic  velocity  (Figure \ref{47Tucvelvsdist})  implying  that
these stars  are being accelerated in  a symmetric way  into the outer
regions  of  the  cluster.   \cite{Drukier07} pointed  out  that  this
evaporation is  exacerbated by the collapse  of a GC  core, leading to
greater  numbers  of two-body  interactions.   This  leads to  greater
numbers of stars accelerated to  the outer regions of the cluster, and
beyond, increasing evaporation.  It  is this evaporation scenario that
we  infer for 47  Tuc, from  its velocity  dispersion at  large radii,
adding  to  the  growing  body  of  evidence  that  47  Tuc  is  in  a
core-collapse, or post-core-collapse, phase \cite[e.g.][and references
therein]{Gebhardt95b,Robinson95,Howell00}.

\new{To confirm that our interpretation  is valid, and that we are not
simply describing  a chance phenomenon,  we rebinned the data  to have
$\sim50$ stars  per bin. The velocity dispersion  profile is unaltered
by the  different binning. Furthermore, we changed  the bin boundaries
to  be sure  that this  increasing velocity  dispersion is  real, and,
again, found  no difference in the  overall shape of  the profile.  Of
course, alternative explanations exist for the increase in dispersion.
For example  if GCs form in  a similar fashion to  Ultra Compact Dwarf
galaxies, there may be a large  quantity of DM in the outskirts of the
cluster  as  discussed by  \cite{Baumgardt08}.   However, no  evidence
exists supporting GCs  forming in this manner.}  A  full MCMC analysis
of the outermost, apparently evaporating, stars will be performed in a
subsequent paper.

\subsection{Mass-to-Light Profiles}\label{ML}

In  dynamical systems  such as  elliptical and  dwarf  galaxies, large
quantities of  dark matter are evidenced by  high mass-to-light ratios
(M/L$_{\rm V}\gg$1).  DM  causes higher stellar accelerations, leading
to inherently  higher maximal stellar  velocities, and hence  a larger
velocity dispersion.  Therefore, one  method for determining whether a
pressure-supported object like a GC  is DM dominated is to measure its
M/L$_{\rm  V}$ from its  velocity dispersion.   For our  M/L$_{\rm V}$
profiles   we   have  used   the   surface   brightness  profiles   by
\citet{Trager95}, converted  to solar luminosities  per square parsec,
and our density profiles were calculated using \citep{Dejonghe87}:

\[
\rho(r) = \frac{M_{tot}}{\pi}\frac{r_s^2}{(r_s^2 + r^2)^2}.
\]

\begin{figure}
  \begin{centering}
  \includegraphics[angle=-90,width=0.45\textwidth]{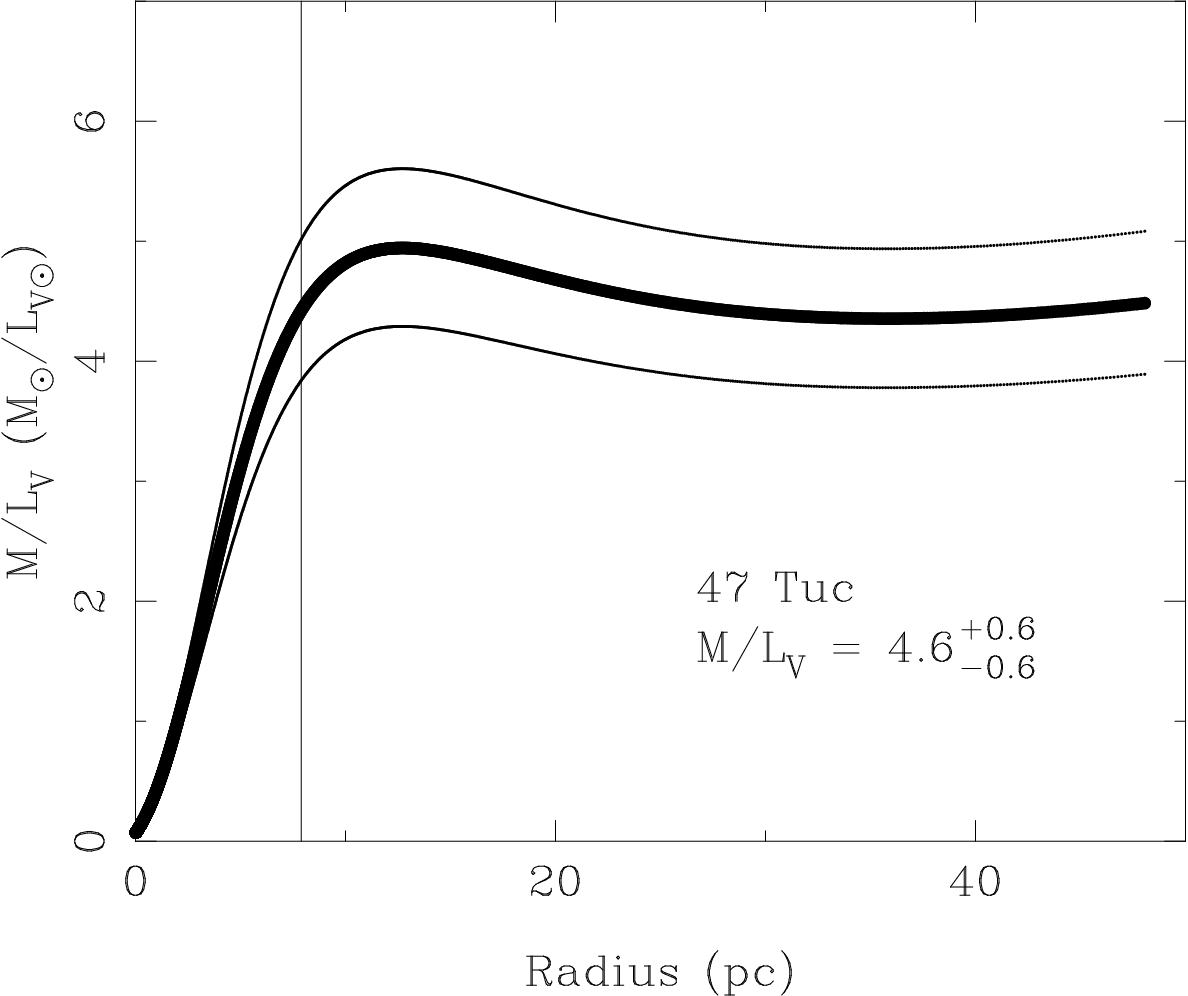}
  \includegraphics[angle=-90,width=0.45\textwidth]{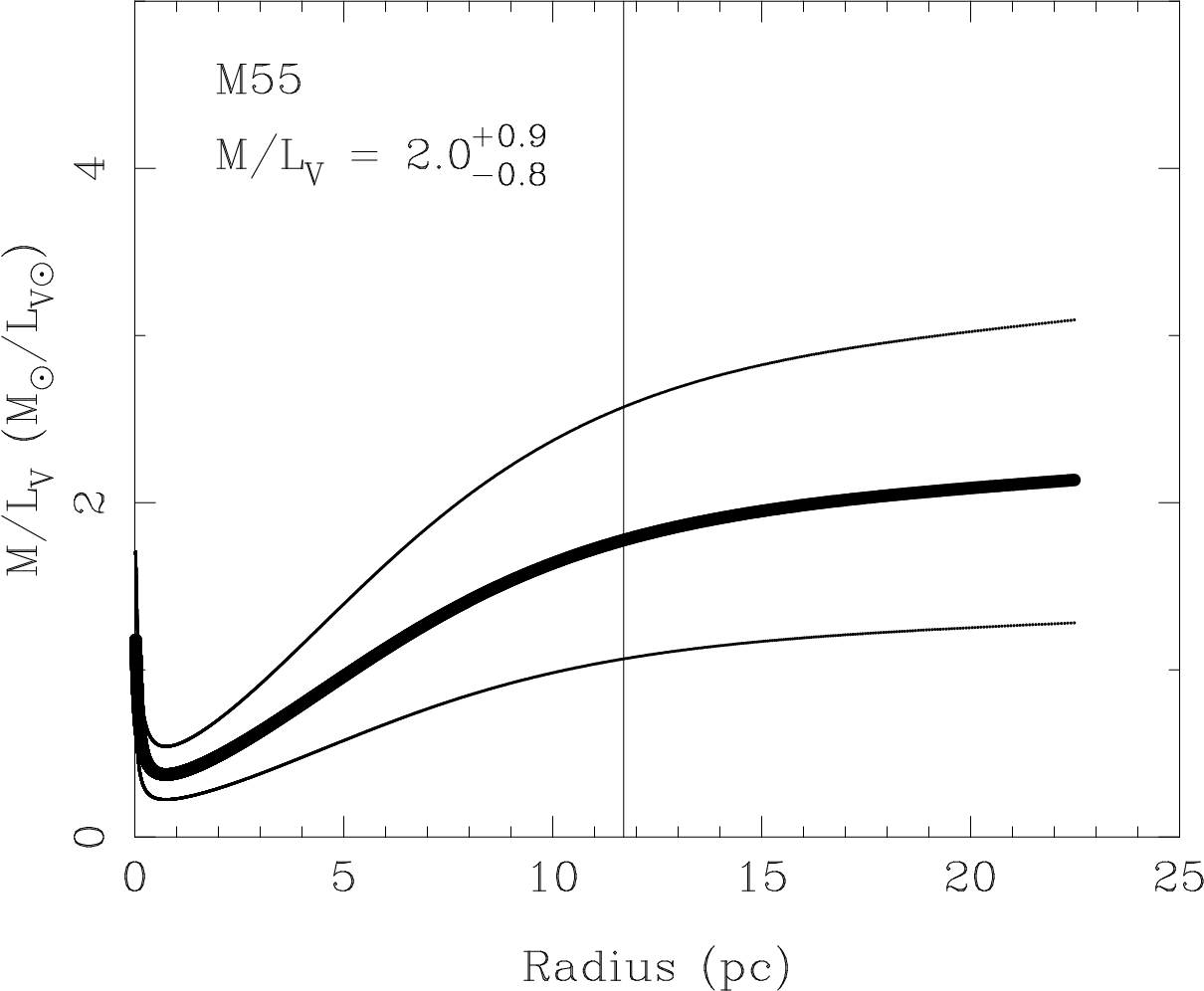}
  \caption{Mass-to-light profiles  of 47 Tuc and  M55.  \new{The thick
  line  is the calculated  M/L$_{\rm V}$  and the  thin lines  are the
  uncertainties. The  vertical line is  $r_s$, and the  mean M/L$_{\rm
  V}$ is  only calculated for $R>r_s$  due to the  uncertainty in core
  mass.}   Neither  cluster  has  M/L$_{\rm V}\gg1$,  indicating  dark
  matter is not dominant.}
  \label{MLfig}
  \end{centering}
\end{figure}

\noindent  The M/L$_{\rm  V}$ profiles,  and associated  M/L$_{\rm V}$
values, are  shown in Figure \ref{MLfig}.  Because  of the uncertainty
in  core  mass,   the  stated  M/L$_{\rm  V}$  values   are  the  mean
mass-to-light for  $R>r_s$.  Our method for  calculating M/L$_{\rm V}$
is preferable to the widely  adopted method using the central mass and
luminosity, due to the uncertainty in core mass.  Despite its apparent
superiority over  other methods for measuring M/L$_{\rm  V}$, very few
studies   have  adopted  it.    This  technique   has  been   used  by
\cite{Gebhardt95a} for 47 Tuc, but only for the inner $10'$ ($\lesssim
r_s$).  Because the mass estimates are highly uncertain at small radii
we do  not claim to have any  realistic data on the  M/L$_{\rm V}$ for
$R<r_s$,  so no  comparison  between  the current  study  and that  by
\cite{Gebhardt95a} can be made.

Neither cluster  has M/L$_{\rm V}\gg1$, therefore  DM cannot dominate,
although the  larger M/L$_{\rm V}$ of  47 Tuc may indicate  a small DM
component.   However,  it  has  been shown  that  ultra-compact  dwarf
galaxies,  which follow  the  same luminosity  -- velocity  dispersion
relation  as old  GCs,  show no  evidence  of DM  for M/L$_{\rm  V}<5$
\cite[e.g.][]{Hasegan05,Evstigneeva07}.   Because of  this, we  see no
requirement for DM in either cluster.

\section{Conclusions}\label{conclusions}

Having the largest sample of spectra ever obtained for 47 Tuc and M55,
we were able to produce a very accurate calibration of [Fe/H] based on
the equivalent  width of  the calcium triplet  lines and the  $K$ band
magnitude  of   the  TRGB.   This   method  is  similar  to   that  by
\cite{Cole04} and \cite{Warren09}, except that we use the TRGB instead of
the  HB which  means this  method can  be used  for much  more distant
objects.

We  calculated  the rotation  of  our  clusters  assuming them  to  be
isothermal.   The  rotation  of  47  Tuc is  $\sim2.2\pm0.2$  with  an
approximate   projected  rotational   axis   along  the   line  PA   =
$40^\circ-220^\circ$,  and  M55  exhibits   rotation  at  a  level  of
$0.25\pm0.09$\,km\,s$^{-1}$  and has an  approximate axis  of rotation
along the  line PA =  $65^\circ-245^\circ$.  For 47 Tuc,  the rotation
amplitude    is    in    good    agreement    with    previous    work
\cite[e.g.][]{Meylan86}.   The  only  previous  study  estimating  the
rotation of  M55 \citep{Szekely07} found  a value about twice  that of
the current work, but \cite{Szekely07} had a sample size approximately
half that of ours, which may explain this discrepancy.

Our calculated velocity dispersion profiles  of 47 Tuc and M55 provide
no evidence that either DM  or a modification of current gravitational
theories  are  required   to  reconcile  their  kinematic  properties,
corroborating previous  work in \citetalias{Lane09}.   The dynamics of
M55 are  well described by a purely  analytic \citet{Plummer11} model,
which  indicates  that  Newtonian  gravity  adequately  describes  its
velocity dispersions,  and shows no breakdown of  Newtonian gravity at
$a_0\approx1.2\times10^{-10}$\,ms$^{-2}$,  as  has  been  claimed  for
other GCs.  The  internal dynamics of 47 Tuc  (for $R<r_t/2$) are also
very  well  described by  the  Plummer  model,  however, the  velocity
dispersion profile of 47 Tuc  exhibits a large increase for $R>r_t/2$,
exactly the  region where evaporation due to  two-body interactions in
the  core   should  be  observable,  especially   for  GCs  undergoing
core-collapse  or in  a post-core-collapse  state.  We  interpret this
increase in  velocity dispersion as  evaporation, and hence  that this
cluster  is either  presently in  a state  of core-collapse,  or  is a
post-core-collapse GC. This  adds to the growing evidence  that 47 Tuc
is  currently  undergoing  a  dynamical phase  change  \cite[e.g.][and
references therein]{Gebhardt95b,Robinson95,Howell00}.  A full analysis
of the  outer regions of  this apparently evaporating cluster  will be
performed in a subsequent paper.

We  used a Plummer  model to  determine the  total mass,  scale radius
($r_s$), and the M/L$_{\rm V}$ profile for each cluster.  We find that
neither  cluster has  M/L$_{\rm  V}\gg1$, indicating  that  DM is  not
dominant.   Within  the uncertainties,  our  estimated cluster  masses
match those  in the  literature well, as  do the M/L$_{\rm  V}$ ratios
\cite[e.g.][]{Meylan89,Pryor93,Meziane96,Kruijssen09}.              The
mass-to-light  profiles  produced   by  \cite{Gebhardt95a}  cannot  be
compared to the current work  because they sampled the inner $10'$ for
which the mass is uncertain. Note that we consider using mass-to-light
{\it profiles} is a more accurate method for calculating M/L$_{\rm V}$
than  using the  core mass  and  surface brightness,  because of  this
uncertainty.

While our results strongly  indicate that the current understanding of
globular clusters being dark matter poor, and their dynamics explained
by  standard Newtonian  gravity,  more robust  dynamical modelling  is
required  for   confirmation.

\section{Acknowledgments}

This  project has  been supported  by  the University  of Sydney,  the
Anglo-Australian  Observatory, the  Australian  Research Council,  the
Hungarian  OTKA  grant K76816  and  the  Lend\"ulet Young  Researchers
Program  of  the Hungarian  Academy  of  Sciences.   We thank  Roberto
Gilmozzi for his  helpful suggestions and feedback on  both this paper
and Paper I.

\bsp

\label{lastpage}


\begin{thebibliography}{99}

\bibitem[Anderson {\it et al.} (2002)]{Anderson02} Anderson, J.~D., 
Laing, P.~A., Lau, E.~L., Liu, A.~S., Nieto, M.~M., 
\& Turyshev, S.~G.\ 2002, Phys. Rev. D, 65, 082004

\bibitem[Angus {\it et al.}(2006)]{Angus06} Angus, G.~W., Famaey, B., 
\& Zhao, H.~S.\ 2006, MNRAS, 371, 138

\bibitem[Balog {\it et al.}(2009)]{Balog09} Balog, Z., Kiss, L.~L., 
Vink{\'o}, J., Rieke, G.~H., Muzerolle, J., G{\'a}sp{\'a}r, A., Young, 
E.~T., \& Gorlova, N.\ 2009, ApJ, 698, 1989

\bibitem[Baumgardt \& Mieske(2008)]{Baumgardt08} Baumgardt, H., \&
  Mieske, S.\ 2008, MNRAS, 391, 942

\bibitem[Bica \& Dutra(2000)]{Bica00} Bica, E., \& Dutra, C.~M.\ 2000, AJ, 119, 1214

\bibitem[Bono {\it et al.}(2008)]{Bono08} Bono, G., et al.\ 2008, 
ApJL, 686, L87

\bibitem[Chou et al.(2007)]{Chou07} Chou, M.-Y., et al.\ 2007, 
ApJ, 670, 346

\bibitem[Cole {\it et al.}(2004)]{Cole04} Cole, A.~A., Smecker-Hane, 
T.~A., Tolstoy, E., Bosler, T.~L., \& Gallagher, J.~S.\ 2004, MNRAS,
347, 367

\bibitem[de Diego(2008)]{deDiego08} de Diego, J.~A.\ 2008, 
Revista Mexicana de Astronomia y Astrofisica Conference Series, 34, 35

\bibitem[Dejonghe(1987)]{Dejonghe87} Dejonghe, H.\ 1987, MNRAS, 
224, 13

\bibitem[Drukier {\it et al.}(2007)]{Drukier07} Drukier, G.~A., Cohn, 
H.~N., Lugger, P.~M., Slavin, S.~D., Berrington, R.~C., 
\& Murphy, B.~W.\ 2007, AJ, 133, 1041

\bibitem[Durrer \& Maartens(2008)]{Durrer08} Durrer, R., \&
  Maartens, R.\ 2008, arXiv:0811.4132

\bibitem[Evstigneeva {\it et al.}(2007)]{Evstigneeva07} Evstigneeva, E.~A., 
Gregg, M.~D., Drinkwater, M.~J., \& Hilker, M.\ 2007, AJ, 133, 1722

\bibitem[Gebhardt \& Fischer(1995)]{Gebhardt95a} Gebhardt, K., \&
  Fischer, P.\ 1995, AJ, 109, 209

\bibitem[Gebhardt {\it et al.}(1995)]{Gebhardt95b} Gebhardt, K., Pryor, 
C., Williams, T.~B., \& Hesser, J.~E.\ 1995, AJ, 110, 1699 

\bibitem[Glatt {\it et al.}(2008a)]{Glatt08a} Glatt, K., et al.\ 2008a, 
AJ, 135, 1106

\bibitem[Glatt {\it et al.}(2008b)]{Glatt08b} Glatt, K., et al.\ 2008b, 
AJ, 136, 1703

\bibitem[Gregory(2005)]{Gregory05} Gregory, P.~C.\ 2005, Bayesian 
Logical Data Analysis for the Physical Sciences: A Comparative Approach 
with `Mathematica' Support. Cambridge University Press, Cambridge,
United Kingdom

\bibitem[Harris(1996)]{Harris96} Harris, W.~E.\ 1996, AJ, 112, 1487

\bibitem[Ha{\c s}egan {\it et al.}(2005)]{Hasegan05} Ha{\c s}egan, M., 
et al.\ 2005, ApJ, 627, 203

\bibitem[Howell {\it et al.}(2000)]{Howell00} Howell, J.~H., 
Guhathakurta, P., \& Gilliland, R.~L.\ 2000, PASP, 112, 1200 

\bibitem[Ibata {\it et al.}(1994)]{Ibata94} Ibata, R.~A., Gilmore, G., \&
  Irwin, M.~J.\ 1994, Nature, 370, 194

\bibitem[Ibata {\it et al.} (2002)]{Ibata02} Ibata, R.~A., Lewis, 
G.~F., Irwin, M.~J., \& Quinn, T.\ 2002, MNRAS, 332, 915

\bibitem[Kiss {\it et al.}(2007)]{Kiss07} Kiss, L.~L., Sz{\'e}kely,
P., Bedding, T.~R., Bakos, G.~{\'A}., \& Lewis, G.~F.\ 2007, ApJL,
659, L129

\bibitem[Kruijssen \& Mieske(2009)]{Kruijssen09} Kruijssen, J.~M.~D.,
  \& Mieske, S.\ 2009, A\&A, 500, 785

\bibitem[Lane  {\it et  al.}(2009)]{Lane09}Lane,  R.~R.  Kiss,  L.~L.,
Lewis,  G.~F.,   Ibata,  R.~A.,   Siebert,  Bedding,  T.~R.,   \&  A.,
Sz{\'e}kely, P.; accepted for publication by MNRAS; (Paper I)

\bibitem[Leon {\it et al.}(2000)]{Leon00} Leon, S., Meylan, G., \& Combes, F.\ 2000, A\&A, 359, 907

\bibitem[Marconi {\it et al.}(1998)]{Marconi98} Marconi, G., Buonanno, R.,
  Castellani, M., Iannicola, G., Molaro, P., Pasquini, L., \& Pulone,
  L.\ 1998, A\&A, 330, 453

\bibitem[Mandushev {\it et  al.} (1991)]{Mandushev91} Mandushev, G., Staneva,
A., \& Spasova, N.\ 1991, A\&A, 252, 94

\bibitem[McLaughlin \& Fall(2008)]{McLaughlin08} McLaughlin, D.~E., \& Fall, S.~M.\ 2008, ApJ, 679, 1272

\bibitem[Meylan \& Mayor(1986)]{Meylan86} Meylan, G., \& Mayor, M.\
  1986, A\&A, 166, 122

\bibitem[Meylan(1989)]{Meylan89} Meylan, G.\ 1989, A\&A, 214, 106

\bibitem[Meziane \& Colin(1996)]{Meziane96} Meziane, K., \& Colin, J.\
  1996, A\&A, 306, 747

\bibitem[Milgrom(1983)]{Milgrom83} Milgrom, M.\ 1983, ApJ, 270, 384

\bibitem[Moffat \& Toth(2008)]{Moffat08} Moffat, J.~W., \&
  Toth, V.~T.\ 2008, ApJ, 680, 1158

\bibitem[Monaco et al.(2004)]{Monaco04} Monaco, L., Pancino, E., 
Ferraro, F.~R., \& Bellazzini, M.\ 2004, MNRAS, 349, 1278

\bibitem[Moore(1996)]{Moore96} Moore, B.\ 1996, ApJL, 461, L13

\bibitem[\protect\citeauthoryear{Munari {\it  et al.}}{2005}]{Munari05}
  {Munari}, U., {Sordo}, R., {Castelli}, F. and {Zwitter}, T. 2005,
  A\&A, 442, 1127
  
\bibitem[Navarro {\it et al.} (1997)]{Navarro97} Navarro, J.~F., Frenk, 
C.~S., \& White, S.~D.~M.\ 1997, ApJ, 490, 493

\bibitem[Odenkirchen {\it et al.} (2001)]{Odenkirchen01} Odenkirchen, M., et 
al.\ 2001, ApJL, 548, L165

\bibitem[Odenkirchen {\it et al.}(2003)]{Odenkirchen03} Odenkirchen, M., et 
al.\ 2003, AJ, 126, 2385

\bibitem[Phinney(1993)]{Phinney93} Phinney, E.~S.\ 1993, Structure and
  Dynamics of Globular Clusters, 50, 141

\bibitem[Plummer(1911)]{Plummer11} Plummer, H.~C.\ 1911, MNRAS, 
71, 460

\bibitem[Pryor \& Meylan(1993)]{Pryor93} Pryor, C., \& Meylan, G.\
  1993, Structure and Dynamics of Globular Clusters, 50, 357

\bibitem[Robinson {\it et al.}(1995)]{Robinson95} Robinson, C., Lyne,
  A.~G., Manchester, R.~N., Bailes, M., D'Amico, N., \& Johnston, S.\
  1995, MNRAS, 274, 547

\bibitem[Sanders \& Noordermeer(2007)]{Sanders07} Sanders, R.~H., \&
  Noordermeer, E.\ 2007, MNRAS, 379, 702

\bibitem[Scarpa {\it et al.}(2003)]{Scarpa03} Scarpa, R., Marconi, G., \& Gilmozzi, R.\ 2003, A\&AL, 405, L15

\bibitem[Scarpa {\it et al.}(2004a)]{Scarpa04a} Scarpa, R., Marconi, G.,
  \& Gilmozzi, R.\ 2004a, Dark Matter in Galaxies, 220, 215

\bibitem[Scarpa {\it et al.}(2004b)]{Scarpa04b} Scarpa R., Marconi
  G. and Gilmozzi R. 2004b, in ``Baryons in Dark Matter Halos'',
  eds. R. Dettmar {\it et al.}, SISSA, Proceedings of Science, 55.1,
http://pos.sissa.it

\bibitem[\protect\citeauthoryear{Scarpa {\it et al.}}{2007}]{Scarpa07}
  {Scarpa}, R., {Marconi}, G., {Gilmozzi}, R. and {Carraro}, G. 2007,
  The Messenger, 128, 41

\bibitem[Skrutskie {\it et al.}(2006)]{Skrutskie06} Skrutskie,
  M.~F., et  al.\ 2006, AJ, 131, 1163

\bibitem[Sollima {\it et al.}(2009)]{Sollima09} Sollima, A., 
Bellazzini, M., Smart, R.~L., Correnti, M., Pancino, E., Ferraro, F.~R., 
\& Romano, D.\ 2009, MNRAS, 396, 2183

\bibitem[\protect\citeauthoryear{Steinmetz {\it et
      al.}}{2006}]{Steinmetz06} {Steinmetz}, M., {\it et al.} 2006,
  AJ, 132, 1645

\bibitem[Strugatskaya(1988)]{Strugatskaya88}    Strugatskaya,   A.~A.\
1988, Pis ma Astronomicheskii Zhurnal, 14, 31

\bibitem[Sz{\'e}kely {\it et al.}(2007)]{Szekely07} Sz{\'e}kely, P., 
Kiss, L.~L., Szatm{\'a}ry, K., Cs{\'a}k, B., Bakos, G.~{\'A}., 
\& Bedding, T.~R.\ 2007, Astronomische Nachrichten, 328, 879

\bibitem[Tabur {\it et al.}(2009)]{Tabur09} Tabur, V., Kiss, L.~L., 
\& Bedding, T.~R.\ 2009, arXiv:0908.2873

\bibitem[Trager {\it et al.}(1995)]{Trager95} Trager, S.~C., King, 
I.~R., \& Djorgovski, S.\ 1995, AJ, 109, 218

\bibitem[Valenti {\it et al.}(2004)]{Valenti04} Valenti, E.,
  Ferraro, F.~R., \& Origlia, L.\ 2004, MNRAS, 354, 815

\bibitem[Warren \& Cole(2009)]{Warren09} Warren, S.~R., \& Cole,
  A.~A.\ 2009, MNRAS, 393, 272

\bibitem[\protect\citeauthoryear{{Zwitter} {\it et al.}}{2008}]{Zwitter08}
  {Zwitter}, T. {\it et  al.} 2008, AJ, 136, 421

\end{thebibliography}
\end{document}